\documentclass[journal]{IEEEtran}
\pdfoutput=1
\usepackage[T1]{fontenc}
\usepackage[cmex10]{amsmath}
\usepackage[cmintegrals]{newtxmath}
\usepackage{commath,bm}
\usepackage[pdftex]{graphicx}
\usepackage{epstopdf}
\usepackage{ifthen}
\usepackage{esint}
\usepackage{cite} 
\usepackage{hyperref}
\usepackage{multirow}
\usepackage{xcolor}
\usepackage{balance}
\newtheorem{theorem}{Theorem}
\newtheorem{corollary}{Corollary}
\newtheorem{remark}{Remark}
\newtheorem{lemma}{Lemma}
\newtheorem{proposition}{Proposition}

\newcommand{\pr}{\text{Pr}}

\newcommand{\Pe}{\mathrm{P_e}}

\newcommand{\erfc}{\mathrm{erfc}\,}

\newcommand{\seq}{\mathbf{b}}

\newcommand{\bit}{b}
\newcommand{\diffd}[1]{D_{#1}} 
\newcommand{\TX}{\text{tx}}
\newcommand{\RX}{\text{rx}}
\newcommand{\vp}[1]{v_{#1}^{\prime}}
\newcommand{\vs}[1]{v_{#1}^{\star}}
\newcommand{\vecp}{\vec{v}^{\prime}}
\newcommand{\vecs}{\vec{v}^{\star}}
\newcommand{\hs}{\hspace{-0.8 mm}}

\allowdisplaybreaks

\begin{document}

\title{Stochastic Channel Modeling for Diffusive Mobile Molecular Communication Systems}
\author{Arman~Ahmadzadeh,~\IEEEmembership{Student Member,~IEEE,}
		Vahid~Jamali,~\IEEEmembership{Student Member,~IEEE,}
		and~Robert~Schober,~\IEEEmembership{Fellow,~IEEE}
\thanks{Manuscript received Sep. 15, 2017; revised Feb. 26, 2018 and May 14, 2018; accepted Jun. 27, 2018. This work was supported in part by the German Science Foundations (Project SCHO 831/7-1), the Friedrich-Alexander University Erlangen-Nuremberg under the Emerging Fields Initiative (EFI), and STAEDTLER Stiftung. This work was presented in part at IEEE GLOBECOM 2017 \cite{Arman4}. The associate editor coordinating the review of this manuscript and approving it for publication was Dr. M. Pierobon. \textit{(Corresponding author: Arman Ahmadzadeh.)}} 
\thanks{A. Ahmadzadeh, V. Jamali, and R. Schober are with the Institute for Digital Communications, University of Erlangen-Nuremberg, D-91058 Erlangen, Germany (email: \{arman.ahmadzadeh, vahid.jamali, robert.schober\}@fau.de).}
}

\maketitle 
\begin{abstract} 
In this paper, we consider mobile molecular communication (MC) systems which are expected to find application in several fields including targeted drug delivery and health monitoring. We develop a mathematical framework for modeling the time-variant stochastic channels of diffusive mobile MC systems. In particular, we consider a diffusive mobile MC system consisting of a pair of transmitter and receiver nano-machines suspended in a fluid medium with a uniform bulk flow, where we assume that either the transmitter, or the receiver, or both are mobile and we model the mobility by Brownian motion. The transmitter and receiver nano-machines exchange information via diffusive signaling molecules. Due to the random movements of the transmitter and receiver nano-machines, the statistics of the channel impulse response (CIR) change over time. We derive closed-form expressions for the mean, the autocorrelation function (ACF), the cumulative distribution function (CDF), and the probability density function (PDF) of the time-variant CIR. Exploiting the ACF, we define the coherence time of the time-variant MC channel as a metric for characterization of the variations of the CIR. The derived CDF is employed for calculation of the outage probability of the system. We also show that under certain conditions, the PDF of the CIR can be accurately approximated by a Log-normal distribution. Based on this approximation, we derive a simple model for outdated channel state information (CSI). Moreover, we derive an analytical expression for evaluation of the expected error probability of a simple detector for the considered MC system. In order to investigate the impact of CIR decorrelation over time, we compare the performances of a detector with perfect CSI knowledge and a detector with outdated CSI knowledge. The accuracy of the proposed analytical expressions is verified via particle-based simulation of the Brownian motion.          
\end{abstract}

\section{Introduction}
Future synthetic nano-networks are expected to facilitate new revolutionary applications in areas such as biological engineering, healthcare, and environmental engineering \cite{NakanoB, AkyildizReview}. Molecular communication (MC), where molecules are the carriers of information, is one of the most promising candidates for enabling reliable communication between nano-machines in such future nano-networks due to its bio-compatibility, energy efficiency, and abundant use in natural biological systems. 

Some of the envisioned application areas of synthetic MC systems may require the deployment of \emph{mobile} nano-machines. For instance, in targeted drug delivery and intracellular therapy applications, it is envisioned that mobile nano-machines carry drug molecules and release them at the desired site of application, see \cite[Chapter 1]{NakanoB}. As another example, in molecular imaging, a group of mobile bio-nano-machines such as viruses carry green fluorescent proteins (GFPs) to gather information about the environmental conditions from a large area inside a body, see \cite[Chapter 1]{NakanoB}. In water quality control, a group of mobile nano-machines may search for small amounts of toxic chemical substances in water supplies, see \cite{AkyildizReview} and references therein. In these applications, communication among the nano-machines is needed for efficient operation. In order to establish a reliable communication link between nano-machines, knowledge of the channel statistics is necessary \cite{PierobonJ4}. However, for \emph{mobile} nano-machines, these statistics change with time, which makes communication even more challenging. Thus, it is crucial to develop a mathematical framework for characterization of the stochastic behaviour of the channel. Stochastic channel models provide the basis for the design of new modulation, detection, and/or estimation schemes for mobile MC systems. 

In the MC literature, the problem of mobile MC has been considered in \cite{108-Luo2016, 106-Hsu2015, 116-Jamali2016, 109-Qiu2016, 105-Guney2012, 107-Kuscu2014, 104-Nakano2016, Iwasaki, Haselmayr}. However, none of these previous works provided a stochastic framework for the modeling of time-variant channels. In particular, in \cite{108-Luo2016, 106-Hsu2015, 109-Qiu2016, 116-Jamali2016} it is assumed that \emph{only} the receiver or the transmitter node is mobile and the channel impulse response (CIR) either changes slowly over time, e.g. due to the slow movement of the receiver, as in \cite{108-Luo2016}, or it is fixed for a block of symbol intervals and may change slowly from one block to the next; see \cite{106-Hsu2015, 116-Jamali2016}. The authors of \cite{109-Qiu2016} consider a mobile macro-robot as transmitter and show that fast movements of the macro-robot can lead to symbol transpositions. The use of positional-distance codes is proposed to mitigate this problem. In \cite{105-Guney2012} and \cite{107-Kuscu2014}, a \emph{three-dimensional} random walk model is adopted for modeling the mobility of nano-machines, where it is assumed that information is \emph{only} exchanged upon the collision of two nano-machines. In particular, \emph{F\"orster resonance energy transfer} and a \emph{neurospike communication model} are considered for information exchange between two colliding nano-machines in \cite{105-Guney2012} and \cite{107-Kuscu2014}, respectively. The authors of \cite{104-Nakano2016} proposed a leader-follower model for target detection applications in two-dimensional mobile MC systems. Langevin equations are used to describe the mobility of the nano-machines. There, it is assumed that the information molecules do not diffuse; the leader nano-machine releases signaling molecules that stick to the release site and form a path that the follower nano-machine follows. The mathematical modeling of this non-diffusion communication approach between leader and follower nano-machines is further analyzed in \cite{Iwasaki}. In the most recent work \cite{Haselmayr}, a one-dimensional random walk model is adopted for modeling the mobility of a point source transmitter and a fully-absorbing point receiver, and the first hitting time distribution of the released particles is evaluated. In our previous work \cite{ArmanJ3}, we have established the mathematical basis required for analyzing mobile MC systems. We have shown that by appropriately modifying the diffusion coefficient of the signaling molecules, the CIR of a mobile MC system can be obtained from the CIR of the same system with fixed transmitter and receiver.

In this paper, similar to \cite{105-Guney2012, 107-Kuscu2014, 104-Nakano2016, Iwasaki, Haselmayr}, both the transmitter and the receiver nano-machines may be mobile. We consider a three-dimensional diffusion model where both the transmitter and receiver nano-machines are subject to diffusion and uniform flow, and unlike in \cite{105-Guney2012, 107-Kuscu2014, 104-Nakano2016, Iwasaki}, the nano-machines exchange information via diffusive signaling molecules. Furthermore, unlike \cite{ArmanJ3}, we develop a \emph{stochastic} framework for characterizing the time-variant CIR of the mobile MC system. In this work, we do not focus on a particular application scenario of mobile MC systems. Instead, motivated by the wide range of possible application scenarios, we adopt a rather general, yet simple system model that captures the main features of mobile MC systems, i.e., the mobility of the nano-machines and the information exchange via molecules. To the best of the authors' knowledge, a stochastic channel model for mobile MC systems has not been reported, yet. In particular, this paper makes the following contributions:     

\begin{enumerate}
	\item Expanding upon our preliminary work in \cite{Arman4}, we establish a mathematical framework for the characterization of the time-variant CIR of mobile MC systems as a stochastic process, i.e., we introduce a stochastic channel model.   
	\item We derive closed-form analytical expressions for the first-order (mean) and second-order (autocorrelation function) moments of the time-variant CIR of mobile MC systems. Equipped with the autocorrelation function of the CIR, we define the coherence time of the channel as the time during which the CIR does not substantially change. 
	\item We derive a closed-form expression for the cumulative distribution function (CDF) of the CIR. The derived CDF can be employed for calculation of the \emph{outage probability} of the considered system. 
	\item We propose a simple model for the outdated CSI in mobile MC systems. To this end, we first derive a closed-form expression for the probability density function (PDF) of the impulse response of the channel. Subsequently, we show that in a certain regime, the PDF can be accurately approximated by a Log-normal distribution. We quantify the approximation regime and based on the approximated PDF, we derive the proposed model for outdated CSI in mobile MC systems.   
	\item To evaluate the impact of the CIR decorrelation occurring in mobile MC systems on performance, we derive the expected bit error probability of a simple detector for perfect and outdated CSI knowledge, respectively.     
\end{enumerate}
We note that this paper expands the corresponding conference version \cite{Arman4} in the following aspects. First, the stochastic channel model in \cite{Arman4} did not include the impact of flow. Second, the closed-form expressions for the CDF and PDF of the time-variant CIR and the model for outdated CSI were not included in \cite{Arman4}.

The rest of this paper is organized as follows. In Section \ref{Sec.SysMod}, we introduce the system model. In Section \ref{Sec.StoChaMod}, we develop the proposed stochastic channel model. In Section \ref{Sec.CDF-PDF}, we derive closed-form expressions for the mean, the autocorrelation function, the CDF, and the PDF of the time-variant CIR. Then, in Section \ref{Sec.PerAna}, we calculate the expected bit error probability of the considered system for detectors with perfect and outdated CSI knowledge, respectively. Simulation and analytical results are presented in Section \ref{Sec.SimRes}, and conclusion are drawn in Section \ref{Sec.Con}.
               
\section{System Model} 
\label{Sec.SysMod}
\begin{figure*}[!tbp]
	\centering
	\begin{minipage}[t]{0.49\textwidth}
	\centering
	\resizebox{\linewidth}{!}{
	\includegraphics[scale = 0.5]{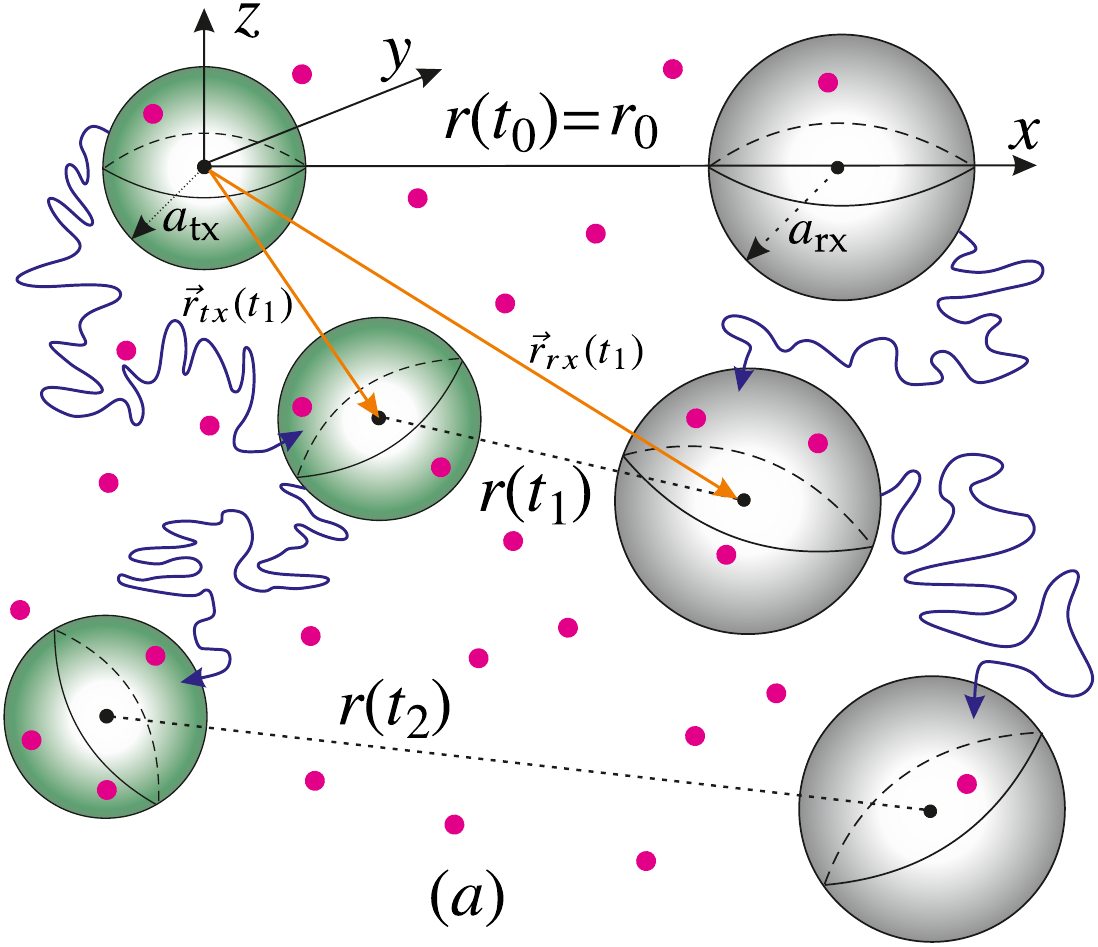}}
	\end{minipage}
	\hfill
  	\begin{minipage}[t]{0.1\textwidth}
  	\end{minipage}
  	\vspace*{-1 mm}
	\begin{minipage}[t]{0.49\textwidth}
	\centering
	\resizebox{\linewidth}{!}{
	\includegraphics[scale = 0.5]{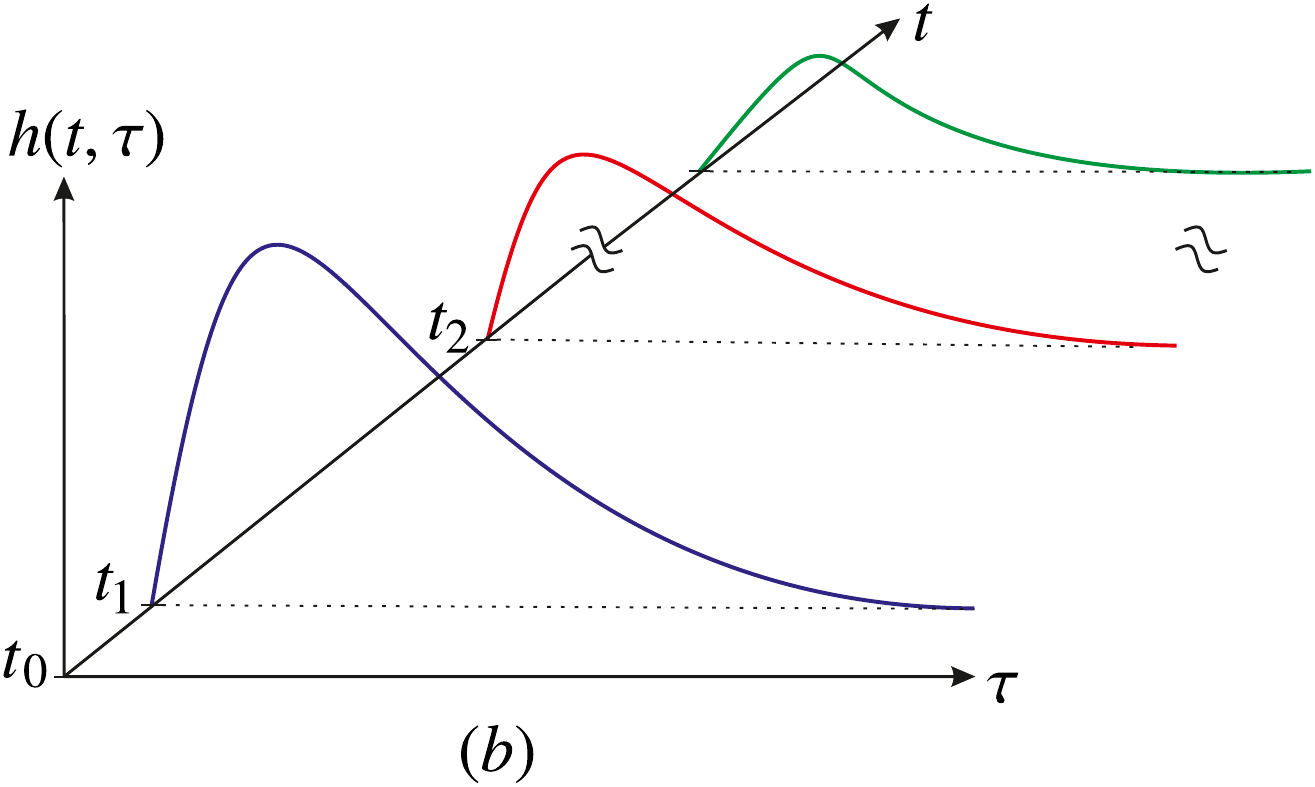}}
	\end{minipage} 
	\caption{(a) Illustration of the considered system model, where the receiver and the 					transmitter are shown as gray and green spheres, respectively. Sample 									trajectories of the receiver and the transmitter are shown as blue solid 								arrows. (b) Example of CIR variation over time $t$.}
	\label{Fig.SystemModel} 
\end{figure*}

We consider an unbounded three-dimensional fluid environment with constant temperature and viscosity. The receiver is modeled as a passive observer\footnote{The model adopted for the receiver nano-machine in this paper can be seen as an abstract and simplified version of more realistic receiver models, where information carrying molecules react with receiver surface receptors, these receptors can be potentially internalized, and the message is transduced via signaling pathways. However, the presented analysis is expected to provide a good approximation for the first-order behaviour of such more elaborate models. In fact, the extension of the derived expressions to more advanced receiver models proposed in the MC literature, see e.g. \cite{ArmanJ2} and \cite{Gohlule}, is an interesting topic for future research. As one example, in Section \ref{Sec.SimRes}, we compare the ACF of the time-variant channel for the passive receiver with that for the reactive receiver model developed in \cite{ArmanJ2}.}, i.e., as a transparent sphere with radius $a_{\RX}$ that diffuses with constant diffusion coefficient $\diffd{\RX}$. For example, small, uncharged molecules, such as ethanol, urea, and oxygen can enter and leave a cell by passive diffusion across the plasma membrane; see \cite[Chapter 16]{AlbertsBook}. Furthermore, we model the transmitter as another transparent sphere with radius $a_{\TX}$ that diffuses with constant diffusion coefficient $\diffd{\TX}$. Here, we adopt a random walk model for transmitter and receiver, since random walk is widely used for modelling the movement of micro-organisms and cells, see \cite{Codling}. The transmitter employs type $A$ molecules, which we also refer to as $A$ molecules and as information or signaling molecules, for conveying information to the receiver. We assume that the $A$ molecules are released in the center of the transmitter and that they can leave the transmitter via free diffusion. We consider a dilute system of $A$ molecules. Hence, we assume that each $A$ molecule diffuses with constant diffusion coefficient $\diffd{A}$ independent of the concentration of the $A$ molecules \cite{SatohBook} and that the diffusion processes of individual $A$ molecules are independent of each other. Moreover, we assume that interfering $A$ molecules are uniformly distributed in the environment and impair the reception. These noise molecules may originate from natural sources in the environment. Furthermore, we assume that there exists a uniform flow in the environment, denoted by $\vec{v} = [v_x, v_y, v_z] $, where $v_x$, $v_y$, and $v_z$ are the components of $\vec{v}$ in the $x$, $y$, and $z$ directions of a Cartesian coordinate system, respectively\footnote{In this work, we consider a biased-random walk model with environmental flow as the source of the bias, where the transmitter, the receiver, and the information molecules can potentially experience flow. The extension of our results to other forms of biased-random walk, where the bias may be caused e.g. by a chemical gradient, is an interesting topic for future research.}. 

Due to the Brownian motion and flow, the positions of the transmitter and the receiver change over time. In particular, we denote the \emph{time-varying} positions of the transmitter and the receiver at time $t$ by $\vec{r}_{\TX}(t)$ and $\vec{r}_{\RX}(t)$, respectively. Then, we define vector $\vec{r}(t) = \vec{r}_{\RX}(t) - \vec{r}_{\TX}(t)$ and denote its magnitude at time $t$ as $r(t)$, i.e., $|\vec{r}(t)| = r(t)$, see Fig.~\ref{Fig.SystemModel}. Furthermore, without loss of generality, we assume that at time $t_0 = 0$, the transmitter is located at the origin of the Cartesian coordinate system, i.e., $\vec{r}_{\TX}(t_0=0) = [0,0,0]$, and the receiver is at $\vec{r}_{\RX}(t_0=0) = [x_0,0,0]$. Thus, $\vec{r}(t_0) = \vec{r}_{\RX}(t_0)$ and $r(t_0 = 0) = r_0 = x_0$.

We assume that the information that is sent from the transmitter to the receiver is encoded into a binary sequence of length $L$, $\seq = [\bit_1, \bit_2, \cdots, \bit_L ]$. Here, $\bit_j$ is the bit transmitted in the $j$th bit interval with $\pr(\bit_j = 1)= P_1$ and $\pr(\bit_j = 0)= P_0 = 1 - P_1$, where $\pr(\cdot)$ denotes probability. We assume that transmitter and receiver are synchronized, see e.g. \cite{JamaliJ4}\footnote{In \cite{JamaliJ4}, symbol synchronization in MC systems is studied. Optimal maximum likelihood synchronization and several practical suboptimal low-complexity synchronization schemes are proposed, see \cite{JamaliJ4} for details.}. We adopt ON/OFF keying for modulation and a fixed bit interval duration of $T$ seconds. In particular, the transmitter releases a fixed number of $A$ molecules, $N_A$, for transmitting bit ``1'' at the \emph{beginning} of a modulation bit interval and no molecules for transmitting bit ``0''.

In this work, we consider simple models for the environmental flow and the receiver nano-machine, and assume perfect symbol synchronization. This allows us to focus on the impact that the \emph{mobility} of the transmitter and receiver nano-machines has on system performance, while keeping the analysis mathematically tractable. Our analytical and simulation results provide first-order insight into the behaviour of mobile MC systems. The extension of the results of this paper to more complex models incorporating e.g. more advanced receiver models \cite{ArmanJ2}, \cite{Gohlule}, and the impact of imperfect symbol synchronization are interesting topics for future research.
\section{Stochastic Channel Model}
\label{Sec.StoChaMod}  
In this section, we provide some preliminaries regarding the modeling of time-variant channels in diffusive mobile MC systems. In particular, in Section \ref{Sec.Time-VariantCIRWithoutFlow}, we introduce the terminology used for describing the time-variant CIR in the absence of flow. Subsequently, we present a mathematical expression for the CIR. Then, we investigate the impact of flow on the derived CIR expression in detail in Section \ref{SubSec.ImpFlow}.
 
\subsection{Impulse Response of Time-Variant MC Channel without Flow} 
\label{Sec.Time-VariantCIRWithoutFlow} 
In this subsection, in order to be able to focus on the impact of the mobility of the transmitter and receiver on the CIR, we assume $\vec{v} = [0, 0, 0]$. We borrow the terminology and the notation for time-variant CIRs from \cite[Ch. 5]{Rappaport}. There, it is assumed that the impulse response of a classical wireless multipath channel can be characterized by a function $h^{\circ}(t,\tau)$, where $t$ represents the time variation due to the mobility of the receiver and $\tau$ describes the channel multipath \emph{delay} for a fixed $t$. Here, we also adopt this notation and derive $h^{\circ}(t,\tau)$ for the problem at hand. In the context of MC, the impulse response of the channel corresponds to the probability of observing a molecule released by the transmitter at the receiver \cite{NakanoB}.  

Let us assume, for the moment, that at the time of release of a given $A$ molecule at the transmitter, $r(t)$ is \emph{known} and given by $r^{\ast}$. Then, the impulse response of the channel, i.e., the probability that a given $A$ molecule, released at the center of the transmitter at time $\tau = 0$, is observed inside the volume of the passive receiver at time $\tau > 0$ can be written as \cite[Eq. (4)]{NoelJ1}
\begin{IEEEeqnarray}{C}
	\label{Eq.CIRConditioned}
	h^{\circ}(\tau|r^{\ast}) = \frac{V_{\mathrm{obs}}}{(4\pi D_1 \tau)^{3/2}} \exp \left( \frac{- (r^{\ast})^2}{4 D_1 \tau} \right),
\end{IEEEeqnarray}
where $V_{\mathrm{obs}} = \frac{4}{3}\pi a_{\RX}^3$ is the volume of the receiver and $D_1 = \diffd{A} + \diffd{\RX}$ is the effective diffusion coefficient capturing the relative motion of the signaling molecules and the receiver, see \cite[Eq.~(8)]{ArmanJ3}. However, due to the random movements of both the transmitter and the receiver, $\vec{r}(t)$ (and consequently $r(t)$) change randomly. In particular, for the problem at hand, the PDF of random variable $\vec{r}(t)$ is given by
\begin{IEEEeqnarray}{C}
	\label{Eq.PDFr(t)NF} 
	\textit{{\large f}}_{\vec{r}(t)}^{\,\,\circ}(\vec{r}) = \frac{1}{(4\pi D_2 t)^{3/2}}\exp \left( \frac{-|\vec{r} - \vec{r}_0|^2}{4 D_2 t} \right),
\end{IEEEeqnarray}
where $D_2 = \diffd{\RX} + \diffd{\TX}$ is the effective diffusion coefficient capturing the relative motion of transmitter and receiver, see \cite[Eq.~(10)]{ArmanJ3}. Thus, for a mobile transmitter and a mobile receiver, the CIR, denoted by $h^{\circ}(t,\tau)$, can be written as 
\begin{IEEEeqnarray}{C}
	\label{Eq.CIRTimVarNF}
	h^{\circ}(t,\tau) = \frac{V_{\mathrm{obs}}}{(4\pi D_1 \tau)^{3/2}} \exp \left( \frac{- |\vec{r}(t)|^2}{4 D_1 \tau} \right),
\end{IEEEeqnarray} 
where $\vec{r}(t)$ is distributed according to the PDF in~\eqref{Eq.PDFr(t)NF}.

CIR $h^{\circ}(t,\tau)$ completely characterizes the time-variant channel and is a function of both $t$ and $\tau$. Variable $t$ represents the time of release of the molecules at the transmitter, whereas $\tau$ represents the relative time of observation of the signaling molecules at the receiver for a fixed value of $t$, cf. Fig.~\ref{Fig.SystemModel}. We note that the movement of the receiver is accounted for in \eqref{Eq.CIRConditioned} via $D_1$ as far as its effect on the $A$ molecules is concerned, and in \eqref{Eq.PDFr(t)NF} via $D_2$ as far as the relative motion of the transmitter and receiver is concerned. Both effects impact $h^{\circ}(t,\tau)$ in \eqref{Eq.CIRTimVarNF}. For any given $\tau$, $h^{\circ}(t,\tau)$ is a stochastic process with random variables $h^{\circ}(t_i, \tau), i \in \{1,2, \ldots, n\}$. Specifically, $h^{\circ}(t_i, \tau)$ can be interpreted as a function of random variable $\vec{r}(t)$.      
\subsection{Impact of Flow}
\label{SubSec.ImpFlow}
In this subsection, we consider the impact of uniform bulk flow on the CIR $h^{\circ}(t,\tau)$. We distinguish between three cases, based on the mobility of the transmitter and the receiver. Let us denote the $x$, $y$, and $z$ coordinates of the position of the transmitter, i.e., the components of vector $\vec{r}_{\TX}(t)$, at time $t$ by $X_{\TX}(t)$, $Y_{\TX}(t)$, and $Z_{\TX}(t)$, respectively. Similarly, the $x$, $y$, and $z$ coordinates of the position of the receiver, i.e., the components of vector $\vec{r}_{\RX}(t)$, at time $t$ are denoted by $X_{\RX}(t)$, $Y_{\RX}(t)$, and $Z_{\RX}(t)$, respectively. 

\subsubsection{Mobile Transmitter and Mobile Receiver}
In this case, both the transmitter and the receiver, along with the information molecules, move with the bulk flow in the environment. Using a moving reference frame that also moves with the bulk flow, it can be easily verified that the expressions for $\textit{{\large f}}_{\vec{r}(t)}^{\,\,\circ}(\vec{r})$ and $h^{\circ}(t, \tau)$, i.e., \eqref{Eq.PDFr(t)NF} and \eqref{Eq.CIRTimVarNF}, respectively, are still valid. However, this is no longer true when one or both of the nano-machines are fixed. These more challenging cases are considered next.

\subsubsection{Mobile Transmitter and Fixed Receiver}
In this case, since the receiver is fixed, $D_1$ and $D_2$ are given by $D_1 = D_A$ and $D_2 = \diffd{\TX}$, respectively. Due to the random walk of the transmitter, and according to the configuration of the transmitter and receiver as shown in Fig.~\ref{Fig.SystemModel}, we can write 
\begin{IEEEeqnarray}{C}
	\label{Eq. XYZDistributionTXRX}
X_{\TX}(t) \sim \mathcal{N}(v_x t,2D_{\TX}t), \,\,\, Y_{\TX}(t) \sim \mathcal{N}(v_y t,2D_{\TX}t),   \nonumber \\
Z_{\TX}(t) \sim \mathcal{N}(v_z t,2D_{\TX}t), \,\,\, X_{\RX}(t) = x_0, \,\,\, Y_{\RX}(t) = 0, \,\,\, Z_{\RX}(t) = 0, \nonumber \\*
\end{IEEEeqnarray}   
where $\mathcal{N}(\mu, \sigma^2)$ denotes a Gaussian distribution with mean $\mu$ and variance $\sigma^2$. Then, for the components of vector $\vec{r}(t) = \vec{r}_{\RX}(t) - \vec{r}_{\TX}(t)$, we obtain
\begin{IEEEeqnarray}{C}
	\label{Eq. XYZDistribution}
	X(t) = X_{\RX}(t) - X_{\TX}(t) \sim \mathcal{N}(x_0 - v_xt, 2D_2t), \nonumber \\
	 Y(t) = Y_{\RX}(t) - Y_{\TX}(t) \sim \mathcal{N}(-v_yt, 2D_2t), \nonumber \\
	Z(t) = Z_{\RX}(t) - Z_{\TX}(t) \sim \mathcal{N}(-v_zt, 2D_2t). 
\end{IEEEeqnarray}

Given \eqref{Eq. XYZDistribution}, the PDF of random variable $\vec{r}(t)$ can be written as  
\begin{IEEEeqnarray}{C}
	\label{Eq.PDFr(t)FixedRX}
	\textit{{\large f}}_{\vec{r}(t)}^{\,\,\RX}(\vec{r}) = \frac{1}{(4\pi \diffd{\TX} t)^{3/2}}\exp \left( \frac{-\big|\vec{r} - \left( \vec{r}_0 - \vec{v}t \right)\big|^2}{4 \diffd{\TX} t} \right).
\end{IEEEeqnarray}
In a similar way, the corresponding impulse response of the time-variant MC channel can be written as 
\begin{IEEEeqnarray}{C}
	\label{Eq.CIRTimVarFixedRX} 
	h^{\RX}(t,\tau) = \frac{V_{\mathrm{obs}}}{(4\pi D_A \tau)^{3/2}} \exp \left( \frac{- \big| \vec{r}(t) - \vec{v}\tau \big|^2}{4 D_A \tau} \right).  
\end{IEEEeqnarray}
The superscript $``\RX "$ in $\textit{{\large f}}_{\vec{r}(t)}^{\,\,\RX}(\vec{r})$ and $h^{\RX}(t,\tau)$ emphasizes that the receiver is fixed. 

\begin{table*}[!t]
	\renewcommand{\arraystretch}{1.5} 
	\centering 
	\caption{Values of $\vec{v}^{\star}$, $\vec{v}^{\prime}$, $\diffd{1}$, and $\diffd{2}$}
	\begin{tabular}{|l||c|c|c|c|}\hline 
		Mobility Scenario & $\vec{v}^{\star}$ & $\vec{v}^{\prime}$ & $\diffd{1}$ & $\diffd{2}$  \\ \hline \hline 
	  No flow with fixed TX and fixed RX & $\vec{0}$ & $\vec{0}$ & $D_A$ & 0 \\ \hline  
      No flow with mobile TX and fixed RX & $\vec{0}$ & $\vec{0}$ & $\diffd{A}$ & $\diffd{\TX}$ \\ \hline
      No flow with fixed TX and mobile RX & $\vec{0}$ & $\vec{0}$ & $D_A + \diffd{\RX}$ & $\diffd{\RX}$ \\ \hline
      No flow with mobile TX and mobile RX & $\vec{0}$ & $\vec{0}$ & $D_A + \diffd{\RX}$ & $\diffd{\RX} + \diffd{\TX}$ \\ \hline 
      Flow with fixed TX and fixed RX & $\vec{0}$ & $\vec{v}$ & $\diffd{A}$ & $0$ \\ \hline
      Flow with mobile TX and fixed RX & $\vec{v}$ & $\vec{v}$ & $\diffd{A}$ & $\diffd{\TX}$ \\ \hline
      Flow with fixed TX and mobile RX & $-\vec{v}$ & $\vec{0}$ & $D_A + \diffd{\RX}$ & $\diffd{\RX}$ \\ \hline        
      Flow with mobile TX and mobile RX & $\vec{0}$ & $\vec{0}$ & $D_A + \diffd{\RX}$ & $\diffd{\RX} + \diffd{\TX}$ \\ \hline 
      \end{tabular}
      \label{Table.2}
\end{table*}

\subsubsection{Fixed Transmitter and Mobile Receiver}
In the third case, the transmitter node is fixed while the receiver is mobile, and, as a result, the corresponding effective diffusion coefficients are given by $D_1 = D_A + \diffd{\RX}$ and $D_2 = \diffd{\RX}$. Now, using a similar approach as in \eqref{Eq. XYZDistributionTXRX} and \eqref{Eq. XYZDistribution}, we can write the PDF of random variable $\vec{r}(t)$ as follows
\begin{IEEEeqnarray}{C}
	\label{Eq.PDFr(t)FixedTX}
	\textit{{\large f}}_{\vec{r}(t)}^{\,\,\TX}(\vec{r}) = \frac{1}{(4\pi \diffd{\RX} t)^{3/2}}\exp \left( \frac{-\big|\vec{r} - \left( \vec{r}_0 + \vec{v}t \right)\big|^2}{4 \diffd{\RX} t} \right),
\end{IEEEeqnarray}
where the superscript $``\TX"$ indicates that the transmitter node is fixed. Furthermore, we denote the impulse response of the time-variant channel for this scenario by $h^{\TX}(t,\tau)$. For calculation of $h^{\TX}(t,\tau)$, however, since both the information molecules and the receiver are equally affected by the flow, by using a moving reference frame, it can be easily verified that $h^{\TX}(t,\tau) = h^{\circ}(t,\tau)$, i.e.,    
\begin{IEEEeqnarray}{C} 
	\label{Eq.CIRTimVarFixedTX} 
	h^{\TX}(t,\tau) = \frac{V_{\mathrm{obs}}}{(4\pi D_1 \tau)^{3/2}} \exp \left( \frac{- | \vec{r}(t)|^2}{4 D_1 \tau} \right). 
\end{IEEEeqnarray} 

By comparing the PDFs of random variable $\vec{r}(t)$ in \eqref{Eq.PDFr(t)NF},  \eqref{Eq.PDFr(t)FixedRX}, and \eqref{Eq.PDFr(t)FixedTX}, and similarly, the impulse responses of the time-variant MC channel in \eqref{Eq.CIRTimVarNF}, \eqref{Eq.CIRTimVarFixedRX}, and \eqref{Eq.CIRTimVarFixedTX}, we can observe that the major difference between the respective equations appears in the argument of the exponential functions. Thus, to facilitate our subsequent analysis, we introduce general expressions for the PDF of $\vec{r}(t)$ and the time-variant CIR unifying all considered cases. In particular, we model the PDF of random variable $\vec{r}(t)$ as 
\begin{IEEEeqnarray}{C}
	\label{Eq.PDFr(t)}
	\textit{{\large f}}_{\vec{r}(t)}(\vec{r}) = \frac{1}{(4\pi \diffd{2} t)^{3/2}}\exp \left( \frac{-\big|\vec{r} - \left( \vec{r}_0 - \vec{v}^{\star}t \right)\big|^2}{4 \diffd{2} t} \right),
\end{IEEEeqnarray}
and the impulse response of the time-variant MC channel as 
\begin{IEEEeqnarray}{C} 
	\label{Eq.CIRTimVar}
	h(t,\tau) = \frac{V_{\mathrm{obs}}}{(4\pi D_1 \tau)^{3/2}} \exp \left( \frac{- \big| \vec{r}(t) - \vec{v}^{\prime}\tau \big|^2}{4 D_1 \tau} \right),  
\end{IEEEeqnarray}
where $\vec{v}^{\star}$, $\vec{v}^{\prime}$, $D_1$, and $D_2$ are defined in Table~\ref{Table.2} for different mobility scenarios. In Table~\ref{Table.2}, ``TX'' and ``RX'' stand for transmitter and receiver, respectively. We also note that for the case of fixed TX and RX (with and without flow), $\textit{{\large f}}_{\vec{r}(t)}(\vec{r}) \to \delta(\vec{r} - \vec{r}_0)$, where $\delta(\cdot)$ is the Dirac delta function. Furthermore, for conciseness of presentation, we introduce the following notations: 
\begin{IEEEeqnarray}{C}
	\label{Eq.Notations}
	\varphi  = \hs \frac{V_{\mathrm{obs}}}{(4 \pi D_1 \tau)^{3/2}}, \lambda(t)  = \frac{1}{(4 \pi D_2 t)^{3/2}},
	\alpha  = \frac{1}{4 D_1 \tau}, \beta(t)  = \frac{1}{4 D_2 t}. \nonumber \\*
\end{IEEEeqnarray}
 
\section{Statistical Analysis of Time-Variant MC Channel} 
\label{Sec.CDF-PDF}
In this section, we first analyze the statistical averages of the considered time-variant channel, i.e., the statistical averages of random process $h(t,\tau)$. In particular, we derive closed-form expressions for the mean and autocorrelation function of $h(t,\tau)$. In addition, we provide an expression for evaluation of the coherence time of the channel. Subsequently, we derive closed-form expressions for the CDF and the PDF of the time-variant CIR, and provide a mathematical model for outdated CSI. 
 
\subsection{Statistical Averages and Coherence Time of Time-Variant MC Channel}
Let us consider first the mean of $h(t,\tau)$ for arbitrary time $t$, denoted by $m(t)$. Then, $m(t)$  can be evaluated as  
\begin{IEEEeqnarray}{C}
	\label{Eq.mean_h_def}
	m(t) = \mathcal{E}\left\lbrace h(t,\tau)\right\rbrace = \hs \hs \int \limits_{\vec{r} \in \mathbb{R}^3} h(t,\tau)\bigg|_{\vec{r}(t) = \vec{r}} \hs \hs \times \textit{{\large f}}_{\vec{r}(t)}(\vec{r}) \dif \vec{r},
\end{IEEEeqnarray}
where $\mathcal{E}(\cdot)$ denotes expectation. The solution to \eqref{Eq.mean_h_def} is provided in the following theorem.
\begin{theorem}[Mean of Time-variant MC Channel] 
The mean of the impulse response of a time-variant MC channel including the effects of uniform bulk flow and diffusive passive transmitter and receiver nano-machines with diffusion coefficients $\diffd{\TX}$ and $\diffd{\RX}$, respectively, which communicate via signaling molecules with diffusion coefficient $\diffd{A}$, is given by
\begin{IEEEeqnarray}{C}
	\label{Eq.mean_h}
\hspace{-4 mm}	m(t) = \frac{V_{\mathrm{obs}}}{\left( 4\pi \left(D_1 \tau + D_2 t \right) \right)^{3/2}} \exp \left( \frac{-|\vec{r}_0 - \vecs t - \vecp \tau|^2}{4 \left(D_1 \tau + D_2 t \right) } \right).
\end{IEEEeqnarray}
\end{theorem}
\begin{IEEEproof} 
Substituting \eqref{Eq.PDFr(t)} and \eqref{Eq.CIRTimVar} in \eqref{Eq.mean_h_def}, we can write $m(t)$ as 
\begin{IEEEeqnarray}{rCl} 
	\label{Eq.mean_h_expand}
	m(t) & = & \varphi \lambda(t) \int \limits_{\vec{r} \in \mathbb{R}^3} e^{-\alpha |\vec{r} - \vec{v}^{\prime}\tau|^2 } \times  e^{ - \beta(t) |\vec{r} - (\vec{r}_0 - \vec{v}^{\star}t)|^2 } \dif \vec{r}. \nonumber \\
	& = & \varphi \lambda(t) \int \limits_{-\infty}^{+\infty} \int \limits_{-\infty}^{+\infty} \int \limits_{-\infty}^{+\infty} e^{ -\left(\alpha + \beta(t)\right)x^2 + 2\left( \beta(t)(x_0 - v_x^{\star}t) + \alpha v_x^{\prime} \tau \right)x} \nonumber \\
	&& \times\> e^{-\left(\alpha + \beta(t)\right)y^2 + 2\left(\alpha v_y^{\prime} \tau -\beta(t)v_y^{\star}t\right)y - \beta(t)(v_y^{\star}t)^2 -\alpha (v_y^{\prime}\tau)^2} \nonumber \\
	&& \times\> e^{ -\left(\alpha + \beta(t)\right)z^2 + 2\left(\alpha v_z^{\prime} \tau -\beta(t)v_z^{\star}t\right)z - \beta(t)(v_z^{\star}t)^2 -\alpha (v_z^{\prime}\tau)^2} \nonumber \\
	&& \times\> e^{- \beta(t)(x_0 - v_x^{\star}t)^2 -\alpha (v_x^{\prime}\tau)^2 } \dif x \dif y \dif z.  
\end{IEEEeqnarray}
The three integrals in \eqref{Eq.mean_h_expand} can be solved independently. Now, using the following definite integral \cite[Eq.~(3.323.2.10)]{Gradshteyn} 
\begin{IEEEeqnarray}{C} 
	\label{Eq.Integral} 
 \int \limits_{-\infty}^{+\infty} \exp \left( -p^2 x^2  \pm qx \right) \dif x = \exp \left(  \frac{q^2}{4p^2}\right) \frac{\sqrt{\pi}}{p},
\end{IEEEeqnarray}
the integrals in \eqref{Eq.mean_h_expand} simplify to the expression in \eqref{Eq.mean_h}. This completes the proof.
\end{IEEEproof} 

\begin{remark}
Since $m(t)$ is a function of $t$, $h(t,\tau)$ is a non-stationary stochastic process. In fact, this is due to the assumption of an unbounded environment, as on average the transmitter and the receiver diffuse away from each other and, ultimately, $h(t,\tau)$ approaches zero as $t \to \infty$.
\end{remark} 

Next, we derive a closed-form expression for the \emph{autocorrelation function} (ACF) of $h(t,\tau)$ for two arbitrary times $t_1$ and $t_2 > t_1$, denoted as $\phi(t_1,t_2)$. To this end, we write $\phi(t_1,t_2)$ as follows~\footnote{In our analysis, the definition of the ACF in \eqref{Eq.ACF_def} can be easily extended to $\phi(t_1,t_2) = \mathcal{E}\left\lbrace h(t_1,\tau_1) h(t_2,\tau_2) \right\rbrace$. However, since in Section~\ref{Sec.PerAna} we consider a detector that takes only one sample at a fixed time after the beginning of each modulation interval, for simplicity of presentation, we focus on the case of $\tau_1 = \tau_2 = \tau$.}
\begin{IEEEeqnarray}{rCl}
	\label{Eq.ACF_def}
	\phi(t_1,t_2) & = & \mathcal{E}\left\lbrace h(t_1,\tau) h(t_2,\tau)\right\rbrace  =  \iint \limits_{\vec{r}_1,\,\vec{r}_2 \in \mathbb{R}^3} h(t_1,\tau)\big|_{\vec{r}(t) = \vec{r}_1} \nonumber \\
	&& \times\> h(t_2,\tau)\big|_{\vec{r}(t) = \vec{r}_2} \times \textit{{\large f}}_{\vec{r}(t_1),\,\vec{r}(t_2)} \left(\vec{r}_1,\, \vec{r}_2 \right) \dif \vec{r}_1 \dif \vec{r}_2, \,\,\, 
\end{IEEEeqnarray} 
where $\textit{{\large f}}_{\vec{r}(t_1),\,\vec{r}(t_2)}\left(\vec{r}_1,\, \vec{r}_2 \right)$ is the joint distribution function of random variables $\vec{r}(t_1)$ and $\vec{r}(t_2)$, which can be written as 
\begin{IEEEeqnarray}{C} 
	\label{Eq.jointPDF}
	\textit{{\large f}}_{\vec{r}(t_1),\,\vec{r}(t_2)}\left(\vec{r}_1,\, \vec{r}_2 \right) = \textit{{\large f}}_{\vec{r}(t_1)}\left( \vec{r}_1 \right) \textit{{\large f}}_{\vec{r}(t_2)} \left( \vec{r}_2 \, \big| \, \vec{r}_1 \right),   
\end{IEEEeqnarray}
where we used the fact that free diffusion is a memoryless process and, as a result, $\textit{{\large f}}_{\vec{r}(t_2)} \left( \vec{r}_2 \, \big| \, \vec{r}_1,\, \vec{r}_0 \right) = \textit{{\large f}}_{\vec{r}(t_2)} \left( \vec{r}_2 \, \big| \, \vec{r}_1 \right)$. Given \eqref{Eq.jointPDF}, a closed-form expression of $\phi(t_1,t_2)$ is provided in the following theorem.

\begin{theorem}[ACF of Time-variant MC Channel] 
The ACF of the impulse response of a time-variant MC channel including the effects of uniform bulk flow and diffusive passive transmitter and receiver nano-machines with diffusion coefficients $\diffd{\TX}$ and $\diffd{\RX}$, respectively, which communicate via signaling molecules with diffusion coefficient $\diffd{A}$, is given by
\begin{IEEEeqnarray}{C}
	\label{Eq.ACF}
\hspace{-5 mm}	\phi(t_1,t_2) = \frac{(2\pi)^3 \varphi^2 \lambda(t_1) \lambda(t_2 - t_1) \exp \left( \kappa_x + \kappa_y + \kappa_z \right)}{\left(4\left(\alpha \hs + \hs \beta \left(t_1\right) \right) \hs \left( \alpha \hs + \hs \beta \left( t_2 \hs - \hs t_1 \right) \right) \hs + \hs \alpha \beta \left( t_2 \hs - \hs t_1 \right)\right)^{3/2}},
\end{IEEEeqnarray} 
where $t_1$ and $t_2 > t_1$ are two arbitrary times and $\kappa_\zeta$ is defined as $\kappa_\zeta = \frac{G_\zeta}{W}$ where  
\begin{IEEEeqnarray}{rCl}
	\label{Eq.ACF_kappa} 
	G_\zeta & = & -\alpha \big[ \left(2\beta(t_2 - t_1) + \alpha\right)\beta(t_1)(x_0 - \vs{\zeta}t_1)^2 + \beta(t_2 - t_1) \nonumber \\ 
	&& \times\> \left(\alpha + \beta(t_1)\right) \left(\vs{\zeta}(t_2 - t_1)\right)^2 - 2\alpha \beta(t_1) \vp{\zeta}\tau(x_0 - \vs{\zeta}t_1) \nonumber  \\ 
	&& +\>  2\beta(t_1)\beta(t_2 - t_1) \vs{\zeta}(t_2 - t_1)(\vp{\zeta}\tau - x_0 + \vs{\zeta}t_1) \nonumber \\
	&& +\>  \left(\alpha + 2\beta(t_2 - t_1)\right) \beta(t_1)(\vp{\zeta}\tau)^2 -4\beta(t_1)\beta(t_2 - t_1) \nonumber \\
	&& \times\> \vp{\zeta}\tau(x_0 - \vs{\zeta}t_1) \big], \nonumber \\ 
	W & = & \alpha \beta(t_1) + 2\alpha \beta(t_2 - t_1) + \beta(t_1) \beta(t_2 - t_1) + \alpha^2,
\end{IEEEeqnarray}
where $\zeta = \{x,y,z\}$ and $x_0$ is set to zero ($x_0 = 0$) when $\zeta = \{y,z\}$.     
\end{theorem}
\begin{IEEEproof}
Please refer to Appendix~\ref{App.1}.
\end{IEEEproof} 

In the following corollary, we study a special case of $\phi(t_1,t_2)$ where $t_2 \to t_1$, i.e., $\phi(t_1, t_1) = \mathcal{E}\left\lbrace h(t_1,\tau) h(t_1,\tau)\right\rbrace$, since $\phi(t_1,t_1)$ cannot be directly obtained from \eqref{Eq.ACF} by substituting $t_2 = t_1$. 
\begin{corollary}[ACF of Time-variant MC Channel for $t_2 = t_1$] In the limit of $t_2 \to t_1$, the ACF of $h(t, \tau)$, i.e., $\phi(t_1, t_1)$, is given by  
\end{corollary}
\begin{IEEEeqnarray}{C}
	\label{Eq.ACF_equaltime}
	\phi(t_1,t_1) = \frac{V_{\mathrm{obs}}^2 \exp \left( \frac{-|\vec{r}_0 - \vecs t - \vecp \tau|^2}{ 2 \left(D_1 \tau + 2 D_2 t_1 \right)} \right)}{\left( 4 \pi D_1 \tau \right)^{3/2} \left( 4\pi \left(D_1 \tau + 2 D_2 t_1 \right) \right)^{3/2}}. 
\end{IEEEeqnarray}
\begin{IEEEproof}
In the limit of $t_2 \to t_1$, $\phi(t_1,t_2)$ in \eqref{Eq.ACF_def} becomes
\begin{IEEEeqnarray}{C} 
	\label{ACF_equaltime_proof_def}
	\phi(t_1,t_1) = \mathcal{E}\left\lbrace h^2(t_1,\tau) \right\rbrace = \int \limits_{\vec{r}_1 \in \mathbb{R}^3} h^2(t_1,\tau)\bigg|_{\vec{r}(t) = \vec{r}_1} \hs \hs \times \textit{{\large f}}_{\vec{r}(t_1)}(\vec{r}_1) \dif \vec{r}_1. \nonumber \\*
\end{IEEEeqnarray}
Substituting \eqref{Eq.PDFr(t)} and \eqref{Eq.CIRTimVar} in \eqref{ACF_equaltime_proof_def}, leads to 
\begin{IEEEeqnarray}{C} 
	\label{Eq.ACF_equaltime_proof_sub} 
 \hspace{-4 mm}	\phi(t_1,t_1) = \varphi^2 \lambda(t_1) \int \limits_{\vec{r}_1 \in \mathbb{R}^3} e^{ -2 \alpha |\vec{r}_1 - \vecp \tau|^2 } \times  e^{ - \beta(t_1) |\vec{r}_1 - (\vec{r}_0 - \vecs t_1)|^2 } \dif \vec{r}_1. \nonumber \\*
\end{IEEEeqnarray} 
Now, expanding the integrand in \eqref{Eq.ACF_equaltime_proof_sub}, similar to \eqref{Eq.mean_h_expand}, and using \eqref{Eq.Integral}, $\phi(t_1,t_1)$ simplifies to \eqref{Eq.ACF_equaltime}.  
\end{IEEEproof} 
Given \eqref{Eq.ACF_equaltime}, we define the variance of the time-variant MC channel as $\sigma^2(t) = \phi(t,t) - m^2(t)$.  

In the remainder of this subsection, we provide an expression for evaluation of the coherence time of the considered time-variant MC channel. To this end, we first define the normalized ACF of random process $h(t,\tau)$ as follow: 
\begin{IEEEeqnarray}{C}
	\label{Eq.CorrCoef_def}
	\rho(t_1,t_2) \hs = \frac{\mathcal{E}\left\lbrace h(t_1,\tau) h(t_2,\tau)\right\rbrace}{\sqrt{\mathcal{E}\left\lbrace h^2(t_1,\tau)\right\rbrace \mathcal{E}\left\lbrace h^2(t_2,\tau)\right\rbrace}} = \frac{\phi(t_1,t_2)}{\sqrt{\phi(t_1,t_1)\phi(t_2,t_2)}}. \nonumber \\*
\end{IEEEeqnarray}
Now, for time $t_1 = 0$, we define the coherence time of the time-variant MC channel, $T^{\text{c}}$, as the minimum time $t_2$ after $t_1 = 0$ for which $\rho(t_1,t_2)$ falls below a certain threshold value $0 < \eta < 1 $, i.e., 
\begin{IEEEeqnarray}{C}
	\label{Eq.CoherenceTime_def}
	T^{\text{c}} = \operatorname*{arg\,min} \limits_{\forall t_2 > 0} \left( \rho(0,t_2) < \eta \right).
\end{IEEEeqnarray}
We note that the particular choice of $\eta$ is application dependent, as the \emph{coherence time} of the channel refers to the time during which the channel does not change \emph{significantly} and the definition of a significant change may vary from one application scenario to another. For example, typical values of $\eta$ reported in the traditional wireless communications literature span the range from $0.5$ to $1$, \cite{Giannakis,Vicario,Wang}, e.g., smaller values of $\eta$ are often employed for resource allocation problems, while larger values of $\eta$ are used for channel estimation problems. Similarly, for MC systems, we expect that future applications that are more sensitive to CIR variations require larger values of $\eta$, e.g. $ 0.8 < \eta \leq 1$, whereas future applications that are more robust to CIR variations can tolerate smaller values of $\eta$, e.g. $0.5 \leq \eta \leq 0.8 $.

\subsection{CDF of Impulse Response of Time-Variant MC Channel}
Next, we are interested in calculating the CDF of the time-variant CIR $h(t,\tau)$ in \eqref{Eq.CIRTimVar}, denoted as $\textit{{\large F}}_{h(t,\tau)}(h)$. Thus, we need to calculate $\pr \left( h(t,\tau) \leq h \right)$. The result of this calculation is provided in the following theorem.

\begin{theorem}[CDF of Time-variant MC Channel]
The CDF of the impulse response of a time-variant MC channel including the effects of uniform bulk flow and diffusive passive transmitter and receiver nano-machines with diffusion coefficients $\diffd{\TX}$ and $\diffd{\RX}$, respectively, which communicate via signaling molecules with diffusion coefficient $\diffd{A}$, is given by 
\begin{IEEEeqnarray}{rCl}
	\label{Eq.CDFofCIRSimplified}
	\textit{{\large F}}_{h(t,\tau)}(h) & = & \frac{\sqrt{D_2t}}{r^{\text{eq}}(t)\sqrt{\pi}} \left\lbrace \exp \left(- \frac{\left( \sqrt{\ln \left( \frac{\varphi}{h} \right)} - r^{\text{eq}}(t) \sqrt{\alpha}\right)^2}{4D_2t\alpha} \right) \right. \nonumber \\
	&& \left. -\> \exp \left(- \frac{\left( \sqrt{\ln \left( \frac{\varphi}{h} \right)} + r^{\text{eq}}(t) \sqrt{\alpha}\right)^2}{4D_2t\alpha} \right) \right\rbrace \nonumber \\ 
	&& +\> \frac{1}{2} \erfc \left( \frac{\sqrt{\ln \left( \frac{\varphi}{h} \right)} + r^{\text{eq}}(t) \sqrt{\alpha}}{\sqrt{4D_2t\alpha}} \right) \nonumber \\ 
	&& +\> \frac{1}{2} \erfc \left( \frac{\sqrt{\ln \left( \frac{\varphi}{h} \right)} - r^{\text{eq}}(t) \sqrt{\alpha}}{\sqrt{4D_2t\alpha}} \right),
\end{IEEEeqnarray}
where $\erfc(\cdot)$ denotes the complementary error function, and we define the \emph{equivalent} distance $r^{\text{eq}}(t) = |\vec{r}_0 - \vecs t - \vecp \tau|$ for compactness.
\end{theorem} 
\begin{IEEEproof}
Please refer to Appendix~\ref{App.2}.
\end{IEEEproof}

\begin{remark}
One immediate application of the derived CDF is the calculation of the outage probability of the considered system. In particular, the outage probability, $P_{\text{out}}$, can be defined as $\pr \left(  h(t,\tau) < h_{\text{min}}\right)$, i.e., the probability that the value of the impulse response of the time-variant channel falls below a minimum threshold. Different criteria can be used for selecting $h_{\text{min}}$, e.g., $h_{\text{min}}$ can be chosen such that it guarantees a minimum bit error probability at the receiver nano-machine. As another application, the derived CDF can be employed for calculation of the average number of successfully transmitted information bits before an outage occurs, denoted by $\bar{n}_b^{\mathrm{out}}$. Let us define $t_{\mathrm{max}} = \operatorname*{arg\,max} \limits_{\forall t > 0} (\pr \left(  h(t,\tau) < h_{\text{min}}\right))$. Then, $\bar{n}_b^{\mathrm{out}} = t_{\mathrm{max}}/T$, where $T$ is the duration of the modulation bit interval.  
\end{remark}

\begin{remark}
Assuming independent diffusion for each information molecule $A$, we can write the observed signal at the receiver as $N(t,\tau) = N_A h(t,\tau)$. Now, given $\textit{{\large F}}_{h(t,\tau)}(h)$, the CDF of $N(t,\tau)$ can be evaluated as $\textit{{\large F}}_{N(t,\tau)}(n) = \textit{{\large F}}_{h(t,\tau)}(n/N_A)$.   
\end{remark} 
\subsection{PDF of Impulse Response of Time-Variant MC Channel}
In this subsection, we calculate the PDF of the impulse response of the time-variant MC channel, and provide a corresponding simple approximation.

\begin{corollary}[PDF of Time-variant MC Channel]
Given \eqref{Eq.CDFofCIRSimplified}, the PDF of the impulse response of the considered time-variant MC channel, $\textit{{\large f}}_{h(t,\tau)}(h)$, can be expressed as 
\begin{IEEEeqnarray}{rCl} 
	\label{Eq.PDFofCIR} 
	\textit{{\large f}}_{h(t,\tau)}(h)
	& = & \hs  \frac{1}{4 \alpha r^{\text{eq}}(t) h \sqrt{\pi D_2 t}} \left[ \exp \left( - \frac{\left( \sqrt{\ln \left( \frac{\varphi}{h} \right)} \hs - r^{\text{eq}}(t) \sqrt{\alpha}\right)^2}{4D_2t\alpha} \right) \right. \nonumber \\
	&& \left. -\> \exp \left( - \frac{\left( \sqrt{\ln \left( \frac{\varphi}{h} \right)} + r^{\text{eq}}(t) \sqrt{\alpha}\right)^2}{4D_2t\alpha} \right) \right].	
\end{IEEEeqnarray}
\end{corollary}
\begin{IEEEproof} 
$\textit{{\large f}}_{h(t,\tau)}(h)$ can be straightforwardly calculated by taking the partial derivative of $\textit{{\large F}}_{h(t,\tau)}(h)$ in \eqref{Eq.CDFofCIRSimplified} with respect to $h$.
\end{IEEEproof}
 
The derived expression for the PDF of the time-variant CIR can be used for the design of new detection and/or estimation algorithms at the receiver nano-machines \cite{JamaliJ3, JamaliL1}. However, \eqref{Eq.PDFofCIR} might be too complicated for some design and/or analysis problems. In the remainder of this section, we first show how \eqref{Eq.PDFofCIR} can be approximated by a Log-normal distribution. Then, we specify the necessary condition that has to be met for this approximation to be accurate. To this end, we start with the following Lemma.

\begin{lemma}
It has been shown in \cite[Ch.~1]{Muirhead} that if random variable $U$ is noncentral chi-squared distributed, i.e., $U \sim \chi_{k}^2(\gamma)$, the asymptotic distribution of 
\begin{IEEEeqnarray}{C}
	\label{Eq.AsyChiSqDef} 
	 \frac{U - (k + \gamma)}{\sqrt{2k + 4\gamma}} \sim \mathcal{N}(0,1) 
\end{IEEEeqnarray}
follows a standard Normal distribution as either $k \to \infty$ for a fixed $\gamma$, or $\gamma \to \infty$ for a fixed $k$.
\end{lemma}

Given the result of Lemma 1, we provide the asymptotic distribution of $h(t,\tau)$ in the following proposition. 
\begin{proposition}
The asymptotic PDF of $h(t,\tau)$ in \eqref{Eq.PDFofCIR}, denoted by $\textit{{\large f}}_{h(t,\tau)}^{\,\,\star}(h)$, in the regime of $\frac{(r^{\text{eq}}(t))^2}{2D_2t} \to \infty$ follows a Log-normal distribution, i.e., 
\begin{IEEEeqnarray}{rCl}
	\label{Eq.PDFofCIRLognormal} 
	\textit{{\large f}}_{h(t,\tau)}^{\,\,\star}(h) & \sim & \text{Lognormal} \left( \mu^{\star}, \sigma^{\star^{2}} \right), \nonumber \\
	\mu^{\star} & = & -2D_2t\alpha \left( 3+\frac{(r^{\text{eq}}(t))^2}{2D_2t}\right) + \ln \left( \varphi \right), \nonumber \\
	\sigma^{\star^{2}} & = & \left( 2D_2t\alpha \right)^2 \times \left( 6 + \frac{2(r^{\text{eq}}(t))^2}{D_2t} \right). 
\end{IEEEeqnarray}
\end{proposition}
\begin{IEEEproof}  
We have shown in Appendix~\ref{App.2} that $h(t,\tau) = \varphi \exp\left( - 2D_2t\alpha \tilde{r}^2(t) \right)$, where $\tilde{r}^2(t) \sim \chi_k^2(\gamma(t))$ with $k = 3$ and $\gamma(t) = (r^{\text{eq}}(t))^2 /( 2D_2t)$. Employing Lemma 1 for the case where $k$ is fixed, in the limit of $\gamma(t) \to \infty$, we obtain $\tilde{r}^2(t) \sim \mathcal{N}\left((k+\gamma(t)), 2k + 4\gamma(t)\right)$, i.e., 
\begin{IEEEeqnarray}{C}
	\label{Eq.AsyChiSq}
	\tilde{r}^2(t) \sim \mathcal{N} \left( 3+\frac{(r^{\text{eq}}(t))^2}{2D_2t}\, , 6 + \frac{2(r^{\text{eq}}(t))^2}{D_2t} \right).
\end{IEEEeqnarray}
Now, given \eqref{Eq.AsyChiSq}, it is straightforward to show that $\textit{{\large f}}_{h(t,\tau)}^{\,\,\star}(h)$ follows a Log-normal distribution.  
\end{IEEEproof}

\begin{figure}[!t]
	\centering
	\label{Fig.GaussianApp}
	\includegraphics[scale=0.55]{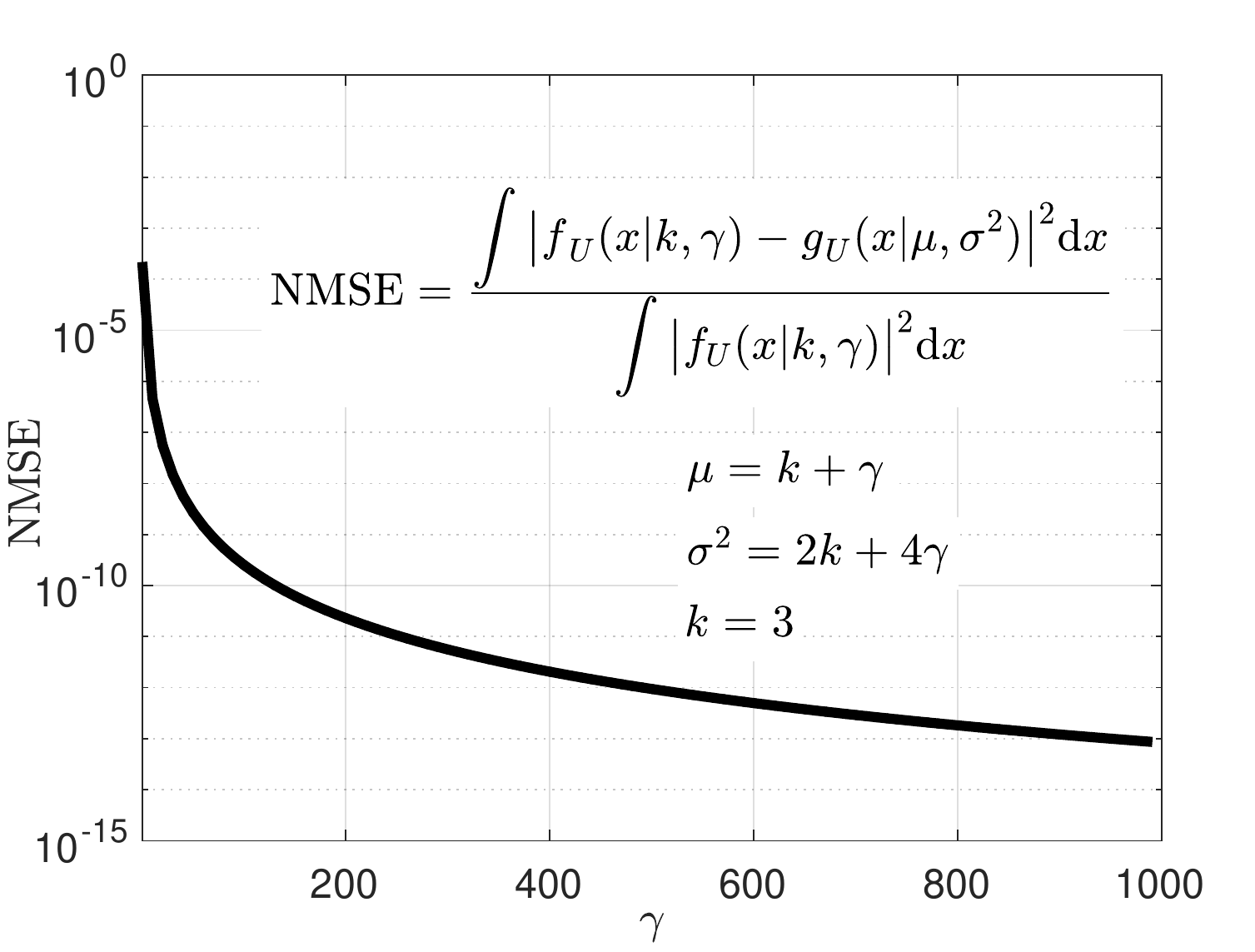}
	\caption{NMSE between the PDF of random variable $U \sim \chi_{k}^2(\gamma)$, $f_U(x|k,\gamma)$, and its Gaussian approximation, $g_U(x|\mu, \sigma^2)$, as a function of $\gamma$.}
\end{figure}

The key step in the derivation of the asymptotic PDF, $\textit{{\large f}}_{h(t,\tau)}^{\,\,\star}(h)$, is the approximation of the PDF of $\tilde{r}^2(t) \sim \chi_k^2(\gamma(t))$ with a Normal distribution employing \eqref{Eq.AsyChiSqDef}. Theoretically, this approximation becomes valid when $\gamma(t) \to \infty$. In order to evaluate the accuracy of the approximation introduced in \eqref{Eq.AsyChiSqDef}, in Fig.~2, we show the normalized mean square error (NMSE) between the PDF of $U \sim \chi_{k}^2(\gamma)$, denoted by $f_U(x|k,\gamma)$, and the approximated PDF of a Gaussian random variable with mean $\mu = (k+\gamma)$ and variance $\sigma^2 = (2k + 4\gamma)$, denoted by $g_U(x|\mu, \sigma^2)$. As can be observed, values of $\gamma \geq 100$ lead to an NMSE of approximately less than $10^{-10}$, which provides an accurate approximation for \eqref{Eq.AsyChiSq}. Taking this into account, we establish a necessary condition for approximating the PDF of the impulse response of a time-variant MC channel by a Log-normal distribution as $\gamma(t) \geq 100$, i.e., 
\begin{IEEEeqnarray}{C}
	\label{Eq.NecCondForLogNormalApp} 
	\frac{(r^{\text{eq}}(t))^2}{2D_2t} \geq 100 \,\,\,\,\,\, \text{or} \,\,\,\,\,\, D_2t \leq \frac{(r^{\text{eq}}(t))^2}{200}. 
\end{IEEEeqnarray}            
\begin{remark}
In the literature of conventional (non-molecular) communication, a similar approach for approximating the distribution of a noncentral chi-squared random variable with a normal distribution can be found. There, usually the case where $k \to \infty$ is considered. For example, in his seminal work \cite{Urkowitz}, Urkowitz showed that values of $k \geq 250$ provide an accurate approximation. Later, Urkowitz' criterion has been widely used in the spectrum sensing literature, see e.g. \cite{Mariani, Tandra, Horgan}. In this work, we have a fixed $k = 3$ and adopt the approximation based on $\gamma \to \infty$.    
\end{remark}

\subsection{Outdated CSI Model}
As one application of the expression derived for the PDF of $h(t,\tau)$, in this subsection, we propose a simple analytical model for outdated CSI in time-variant MC channels. In particular, for the problem at hand, knowledge of the CSI is equivalent to knowledge of the CIR. Due to the mobility of the nano-machines, the CIR decorrelates over time, which may limit the performance of detection algorithms that require instantaneous knowledge of the CIR. Similar to conventional wireless communication systems, one possible approach would be to organize the transmitted symbols into blocks, estimate the CIR at the beginning of each block based on pilot symbols, and use the estimated CIR for detection/decoding of the symbols in the block, where the CIR changes within the transmission block due to the mobility of transmitter and receiver. On the other hand, there is a trade-off between block length and CSI quality, i.e., by increasing the block length, the CSI becomes more outdated but the training overhead is reduced. Thus, a simple yet accurate model for the outdated CSI is desirable.  

Let us assume that the receiver obtains a \emph{perfect} estimate of the CIR at time $t_s$, where we denote the estimated CIR by $\hat{h}(t_s,\tau)$ and the estimated $\vec{r}(t_s)$ by $\hat{\vec{r}}_s$.\footnote{We note that based on the expression for the CIR in \eqref{Eq.CIRTimVar}, knowing $h(t_s,\tau)$ is equivalent to knowing $\vec{r}(t_s)$, if all other system parameters are known at the receiver.} Now, given \eqref{Eq.AsyChiSq}, for $t > t_s$, we can write
\begin{IEEEeqnarray}{rCl}
	\label{Eq.OutDatedCSI_Model}
	h(t^{\prime},\tau) & = & \varphi \exp\left( - 2D_2 t^{\prime} \alpha \tilde{r}^2(t^{\prime}) \right) \nonumber \\
	 & = &  \varphi \exp\Bigg( - 2D_2t^{\prime}\alpha \bigg( 3+\frac{(r^{\text{eq}}(t^{\prime}))^2}{2D_2t^{\prime}} \nonumber \\
	 && +\> \sqrt{6 + \frac{2(r^{\text{eq}}(t^{\prime}))^2}{D_2t^{\prime}}} \times \epsilon \bigg) \Bigg),
\end{IEEEeqnarray} 
where $t^{\prime} = t - t_s$ and $\epsilon \sim \mathcal{N}(0,1)$. Now, substituting $r^{\mathrm{eq}}(t^{\prime}) = \sqrt{|\hat{\vec{r}}_s - \vecs t^{\prime} - \vecp \tau|^2}$ into \eqref{Eq.OutDatedCSI_Model}, it can be easily verified that $h(t^{\prime},\tau)$ can be written as
\begin{IEEEeqnarray}{C}
	\label{Eq.OutDatedCSI}
	h(t^{\prime},\tau) = C \hat{h}(t_s,\tau) M^{\Theta},  
\end{IEEEeqnarray} 
where $M \sim \mathrm{Lognormal}(0,1)$, and $C$ and $\Theta$ are defined as 
\begin{IEEEeqnarray}{rCl}
	\label{Eq.OutDatedCSI2}
	C & = & \exp \left( - 6D_2t^{\prime}\alpha -2 \alpha \left( \vecs t^{\prime} \odot (\hat{\vec{r}}_s - \vecp \tau) \right) \right),\nonumber \\ 
	 \Theta & = & - 2D_2t^{\prime}\alpha \sqrt{6 + \frac{2|\hat{\vec{r}}_s - \vecs t^{\prime} - \vecp \tau|^2}{D_2t^{\prime}}},
\end{IEEEeqnarray}
and $\odot$ denotes the inner product of two vectors. In \eqref{Eq.OutDatedCSI}, $C$ and $\Theta$ are two time-dependent variables. In the limit of $t \to t_s$ ($t^{\prime} \to 0$), $C$ and $\Theta$ approach $1$ and $0$, respectively, and $h(t^{\prime},\tau) \to \hat{h}(t_s,\tau)$. On the other hand, as $t^{\prime}$ increases, $C$ decreases and $\Theta$ increases, which reflects the decorrelation of $h(t^{\prime},\tau)$ and $\hat{h}(t_s,\tau)$. Furthermore, we note that the accuracy of \eqref{Eq.OutDatedCSI} depends on the accuracy of the approximation introduced in \eqref{Eq.AsyChiSq}.               
\section{Error Rate Analysis for Perfect and Outdated CSI}
\label{Sec.PerAna}
In this section, we first calculate the expected error probability of a single-sample threshold detector. Then, we discuss the choice of the detection threshold of the detector. Finally, in order to investigate the impact of CIR decorrelation, we calculate the expected error probability of the considered detector for perfect and outdated CSI.

\subsection{Expected Bit Error Probability}
We consider a single-sample threshold detector, where the receiver takes one sample\footnote{In nature, cells measure (count) signaling molecule via receptor protein molecules covering their surface. These measurements are inherently random due to several noise sources such as a) the stochastic random walk of the signaling molecules, b) the stochastic nature of the reactions occurring in the channel, c) the stochastic binding and unbinding of the signaling molecules with the receptor protein molecules, and d) the stochastic nature of the signaling pathways relaying the receptors' signals into the cell, see e.g. [34]. In this work, we assume that the receiver counting process is impaired only by noise source a). However, in Section VI, as one example, we also consider the reactive receiver model developed in \cite{ArmanJ2}, where the counting process is impaired by noise sources a), b), and c), and compare the ACFs of the time-variant channel for the passive and the reactive receiver models.} at a fixed time $\tau_s$ after the release of the molecules at the transmitter in each modulation bit interval, counts the number of signaling $A$ molecules inside its volume, and compares this number with a detection threshold. In particular, we denote the received signal, i.e., the number of molecules observed inside the volume of the receiver in the $j$th bit interval, $j \in \{1,2,\ldots, L\}$, at the time of sampling by $N(\tau_{j,s})$, where $\tau_{j,s} = (j-1)T + \tau_s$. Furthermore, we assume that the detection threshold of the receiver, $\xi_j$, can be adapted from one bit interval to the next. The choice of $\xi_j$ is discussed in the next subsection. Thus, the decision of the single-sample detector in the $j$th bit interval, $\hat{\bit}_j$, is given by
\begin{equation}
	\label{Eq.Reception} 
	\hat{\bit}_j = \begin{cases} 
	1 &\mbox{if } N(\tau_{j,s}) \geq \xi_j, \\
	0 &\mbox{if } N(\tau_{j,s}) < \xi_j. 
			\end{cases}
\end{equation}

For the decision rule in \eqref{Eq.Reception}, we showed in \cite{ArmanJ3} that the expected error probability of the $j$th bit, $\mathrm{\overline{P}_e}(b_j)$, can be calculated as \cite[Eq.~(12)]{ArmanJ3}
\begin{IEEEeqnarray}{C}
	\label{Eq.BER_bj}
 \hspace{-6mm}	\mathrm{\overline{P}_e}(b_j) = \idotsint\limits_{\mathbf{r} \in \mathcal{R}} \sum_{\seq \in \mathcal{B}} \textit{{\large f}}_{\vec{R}}\left( \mathbf{r} \right) \pr(\seq) \Pe(\bit_j | \seq, \mathbf{r}) \dif \vec{r}_1 \cdots \dif \vec{r}_{L-1}, 
\end{IEEEeqnarray}
where $\textit{{\large f}}_{\vec{R}}\left( \mathbf{r} \right)$ is the ($L-1$)-dimensional joint PDF of vector $\vec{R} = [\vec{r}(T), \vec{r}(2T), \cdots, \vec{r}((L-1)T)]$ that can be evaluated as      
\begin{IEEEeqnarray}{rCl}
	\label{Eq.JointPDF}
	\textit{{\large f}}_{\vec{R}}\left( \mathbf{r} \right) & = & \textit{{\large f}}_{\vec{r}(T)}\left( \vec{r}_1 | \vec{r}_{0} \right) \times \cdots \times \textit{{\large f}}_{\vec{r}\left((L-1)T\right)}\left( \vec{r}_{L-1} | \vec{r}_{L-2}, \cdots, \vec{r}_0 \right) \nonumber \\
	 & \overset{(a)}{=} & \prod \limits_{j = 1}^{j=L-1} \textit{{\large f}}_{\vec{r}(jT)}\left( \vec{r}_j | \vec{r}_{j-1} \right).
\end{IEEEeqnarray}
Here, $\mathbf{r} = [\vec{r}_1, \vec{r}_2, \cdots, \vec{r}_{L-1}]$ is one sample realization of $\vec{R}$ and equality $(a)$ holds as free diffusion is a memoryless process, i.e., $\textit{{\large f}}_{\vec{r}(jT)}\left( \vec{r}_j | \vec{r}_{j-1}, \cdots, \vec{r}_0 \right) = \textit{{\large f}}_{\vec{r}(jT)}\left( \vec{r}_j | \vec{r}_{j-1} \right)$. Furthermore, $\mathcal{R}$ and $\mathcal{B}$ are the sets containing all possible realizations of $\mathbf{r}$ and $\seq$, respectively, and $\pr(\seq)$ denotes the likelihood of the occurrence of $\seq$ and $\Pe(\bit_j | \seq, \mathbf{r})$ is the conditional bit error probability of $\bit_j$. In \cite{ArmanJ3}, we considered a \emph{reactive receiver} \cite{ArmanJ2} and showed how $\Pe(\bit_j | \seq, \mathbf{r})$ can be calculated for a single-sample detector using a \emph{fixed} detection threshold $\xi$. Here, we provide $\Pe(\bit_j | \seq, \mathbf{r})$ for a \emph{passive receiver} \cite{NoelJ1} employing a single-sample detector with an \emph{adaptive} detection threshold $\xi_j$. 

Let us assume that $\seq$ and $\mathbf{r}$ are known. It has been shown in \cite{NoelJ1} that the number of observed molecules, $N(\tau_{j,s})$, can be accurately approximated by a Poisson random variable. The mean of $N(\tau_{j,s})$, denoted by $\overline{N}(\tau_{j,s})$, due to the transmission of all bits up to the current bit interval can be written as 
\begin{IEEEeqnarray}{C} 
	\label{Eq.MeanReceivedSig} 
\hspace{-4mm}	\overline{N}(\tau_{j,s}) = N_{A} \sum_{i=1}^{j} \bit_i h \left( iT, (j-i)T +\tau_{s} \right)\bigg|_{\vec{r}(iT) = \vec{r}_i} + \overline{n}_A,
\end{IEEEeqnarray} 
where $\overline{n}_A$ is the mean number of noise molecules inside the volume of the receiver at any given time. Now, given $\overline{N}(\tau_{j,s})$ and the decision rule in \eqref{Eq.Reception}, $\Pe(\bit_j | \seq, \mathbf{r})$ can be written as 
\begin{IEEEeqnarray}{C}
	\label{Eq.BER_bj_Conditional}
	\Pe(\bit_j | \seq, \mathbf{r}) = \begin{cases} 
	\pr(N(\tau_{j,s}) < \xi_j) & \mbox{if } \bit_j = 1, \\
	\pr(N(\tau_{j,s}) \geq \xi_j) &\mbox{if } \bit_j = 0, 
			\end{cases} 
\end{IEEEeqnarray}
where $\pr(N(\tau_{j,s}) < \xi_j)$ can be calculated from the cumulative distribution function of a Poisson distribution as 
\begin{IEEEeqnarray}{C} 
	\label{Eq.Poisson_CDF}
	\pr(N(\tau_{j,s}) < \xi_j)  =  \exp\left(-\overline{N}(\tau_{j,s})\right) \sum_{\omega = 0}^{\xi_j - 1} \frac{\left(\overline{N}(\tau_{j,s})\right)^{\omega}}{\omega !},
\end{IEEEeqnarray}
and $\pr(N(\tau_{j,s}) \geq \xi_j) = 1 - \pr(N(\tau_{j,s}) < \xi_j)$. Given $\mathrm{P_e}(\bit_j| \seq, \mathbf{r})$ in \eqref{Eq.BER_bj_Conditional}, $\mathrm{\overline{P}_e}(\bit_j)$ can be calculated based on \eqref{Eq.BER_bj}. Subsequently, we can obtain the expected error probability as $\mathrm{\overline{P}_e} = \frac{1}{L} \sum_{j=1}^{L} \mathrm{\overline{P}_e}(\bit_j)$.   
 
In the remainder of this subsection, we discuss the choice of the adaptive detection threshold for the considered single-sample detector. Let us assume for the moment that sequence $\{\bit_1, \bit_2, \cdots, \bit_{j-1}\}$ and $\mathbf{r}$ are known, and we are interested in finding the optimal detection threshold, $\xi_j^{\text{opt}}$, that minimizes the instantaneous error probability $\mathrm{P_e}(\bit_j)$. Then, we have shown in \cite{ArmanJ1} that for any threshold detector whose received signal can be modeled as a Poisson random variable, $\xi_j^{\text{opt}}$ is given by \cite[Eq.~(25)]{ArmanJ1}
\begin{IEEEeqnarray}{C} 
	\label{Eq.OptimalThreshold}
	\xi_j^{\text{opt}} = \Bigg \lceil \frac{\ln \left( \frac{P_0}{P_1} \right) + \left( \lambda_1 - \lambda_0 \right)}{\ln \left( \lambda_1 / \lambda_0 \right)} \Bigg \rceil,
\end{IEEEeqnarray} 
where $\lambda_1 = \overline{N}(\tau_{j,s}|\bit_j = 1)$,  $\lambda_0 = \overline{N}(\tau_{j,s}|\bit_j = 0)$, and $\lceil \cdot \rceil$ denotes the ceiling function. 
\begin{remark}
We note that the evaluation of $\xi_j^{\text{opt}}$ requires knowledge of the previously transmitted bits up to the current bit interval, which is not available in practice. Thus, for practical implementation, we propose a suboptimal detector whose detection threshold, $\hat{\xi}_j^{\text{subopt}}$, is evaluated according to \eqref{Eq.OptimalThreshold} after replacing $\{ \bit_1, \bit_2, \ldots \bit_{j-1}\}$ in \eqref{Eq.MeanReceivedSig} with the estimated previous bits, i.e., $\{\hat{\bit}_1, \hat{\bit}_2, \cdots, \hat{\bit}_{j-1}\}$.      
\end{remark} 
\begin{remark}
It has been shown in \cite{JamaliC1} that when the effect of inter-symbol interference (ISI) is negligible compared to $\overline{n}_A$, the combination of \eqref{Eq.Reception} and \eqref{Eq.OptimalThreshold} constitutes the optimal maximum likelihood (ML) detector. We note that, in this regime, knowledge of previously transmitted bits is not required for calculation of $\xi_j^{\text{opt}}$.
\end{remark} 
\subsection{Detectors with Perfect and Outdated CSI} 
In this subsection, we distinguish between the cases of \emph{perfect} CSI and \emph{outdated} CSI knowledge, and explain how the corresponding expected error probabilities of the single-sample detector can be evaluated.

\textit{Perfect CSI:} For the case of a single-sample detector with perfect CSI, we assume that for any given modulation bit interval, $r(t)$ is known at the receiver for all previous bit intervals up to the current bit interval, i.e., for the $j$th bit interval, $[\vec{r}(0), \vec{r}(T), \ldots, \vec{r}(jT)]$ is known at the receiver. Thus, $\hat{\xi}_j^{\text{subopt}}$ can be directly obtained from \eqref{Eq.CIRTimVar}, \eqref{Eq.MeanReceivedSig}, and \eqref{Eq.OptimalThreshold}.

\textit{Outdated CSI:} For the case of a single-sample detector with outdated CSI, we assume that \emph{only} the initial distance between transmitter and receiver at time $t_0 = 0$, i.e., $r_0$, is known at the receiver. As a result, in any modulation bit interval, the receiver evaluates $\hat{\xi}_j^{\text{subopt}}$ via \eqref{Eq.OptimalThreshold} with the mean given by
\begin{IEEEeqnarray}{C} 
	\label{Eq.MeanOutDatedCSI} 
\hspace{-4mm}	\overline{N}(\tau_{j,s}) = N_{A} \sum_{i=1}^{j} \bit_i h \left( t_0\,, (j-i)T +\tau_{s} \right) + \overline{n}_A.
\end{IEEEeqnarray}
 
Finally, for both cases, $\Pe(\bit_j | \seq, \mathbf{r})$ is obtained from
\eqref{Eq.BER_bj_Conditional}.           
\section{Simulation Results}
\label{Sec.SimRes}
\begin{table}
	\renewcommand{\arraystretch}{1.5} 
	\centering 
	\caption{Simulation Parameters}
	\begin{tabular}{|c|c|  |c|c|}\hline 
      Parameter & Value & Parameter & Value \\ \hline \hline
        $N_A$ & $30000$ & $T$ & $0.5$ ms  \\ \hline
        $D_A$ & $5 \times 10^{-9}$ $\text{m}^2/{\text{s}}$ & $\tau = \tau_s$ & $0.035$ ms \\ \hline
        $D_{\text{rx}}$ & $10^{-13}$ $\text{m}^2/{\text{s}}$ & $L$   & $50$ \\ \hline
        $r_0$ & $1$\, $\mu$m & $P_1$   & $0.5$\\ \hline
        $a_{\text{rx}}$   & $0.15$\, $\mu$m &  $P_0$ & 0.5  \\ \hline
        $\overline{n}_A$   & $10$ &  $\Delta t$ & 5 $\mu$s \\ \hline
      \end{tabular} 
      \label{Table.1}
\end{table}
In this section, we present simulation and analytical results to assess the accuracy of the derived analytical expressions for the statistics of the time-variant CIR and the expected error probability of the considered mobile MC system. 
For simulation, we developed a particle-based simulator of Brownian motion\footnote{We employ a standard particle-based Brownian motion algorithm \cite{Steven_Andrews}, as this approach, unlike other approaches that are based upon the \emph{reaction-diffusion master equation} \cite{Hattne}, does not rely on mesoscopic lengths and time scales for which the system has to be well stirred.}, where the precise locations of the signaling molecules, transmitter, and receiver are tracked throughout the simulation environment. In particular, in the simulation algorithm, time is advanced in discrete steps of $\Delta t$ seconds. In each step of the simulation, each $A$ molecule, the transmitter, and the receiver undergo random walks, and their new positions in each Cartesian coordinate are obtained by sampling a Gaussian random variable with mean $v_\zeta \Delta t$, $\zeta = \{x,y,z \}$, and standard deviation $\sqrt{2D_A\Delta t}$, $\sqrt{2 \diffd{\TX} \Delta t}$, and $\sqrt{2 \diffd{\RX} \Delta t}$, respectively. Furthermore, we used Monte-Carlo simulation for evaluation of the multi-dimensional integral in \eqref{Eq.BER_bj}.   

For all simulation results, we chose the set of simulation parameters provided in Table~\ref{Table.1}, unless stated otherwise. For all simulation results in Sections \ref{Sec.SimFirstandSecMom} and \ref{Sec.SimCDFPDF}, we assume that $\bar{n}_A = 0$. Furthermore, we considered an environment with the viscosity of water ($\simeq 0.89\,\text{mPa}\cdot\text{s}$) at $25\, \mathrm{{}^{\circ}C}$ and we used the Stokes--Einstein equation \cite[Eq.~(5.7)]{NakanoB} for calculation of $D_A$ and $\diffd{\TX}$. The only parameters that were varied are $\diffd{\TX}=\{0.1, 1, 5, 20, 100\}\times 10^{-13}$ $\frac{\text{m}^2}{\text{s}}$ (corresponding to $a_{\text{tx}} = 2.4537 \times \{10^{-5}, 10^{-6}, 2 \times 10^{-7}, 5 \times 10^{-8}, 10^{-8} \}$ m)\footnote{The very small values of $a_{\text{tx}}$ (in the order of tens of nm) have been used only to be able to consider the full range of $\diffd{\TX}$ values.} and flow velocity $\vec{v}$\footnote{Example environments, where parameter values similar to those assumed in this section occur, include 1) micro-fluidic channels \cite{Berthier}, and 2) cytoplasmic streaming, see e.g. \cite{Goldstein}, \cite{Verchot}. Typical values of $\vec{v}$ in micro-fluidic channels are in the range of a few microns per second to a few millimetres per second. In cytoplasmic streaming, the intracellular flow originates from the motion of the motor protein myosin along filamentary actin strands, and depending on the cell size, the flow velocities range from a few microns per second to tens of microns per second. Here, we adopt flow velocities of a few microns per second, in order to be able to show the impact of flow on the performance of mobile MC systems over the time scale that is simulated. For application scenarios where flow does not exist, we have $\vec{v} = [0,0,0]$.}. All simulation results were averaged over $10^{5}$ independent realizations of the environment.  

\subsection{First- and Second-order Moments of CIR}
\label{Sec.SimFirstandSecMom}
\begin{figure}[!t]
	\centering
	\includegraphics[scale=0.55]{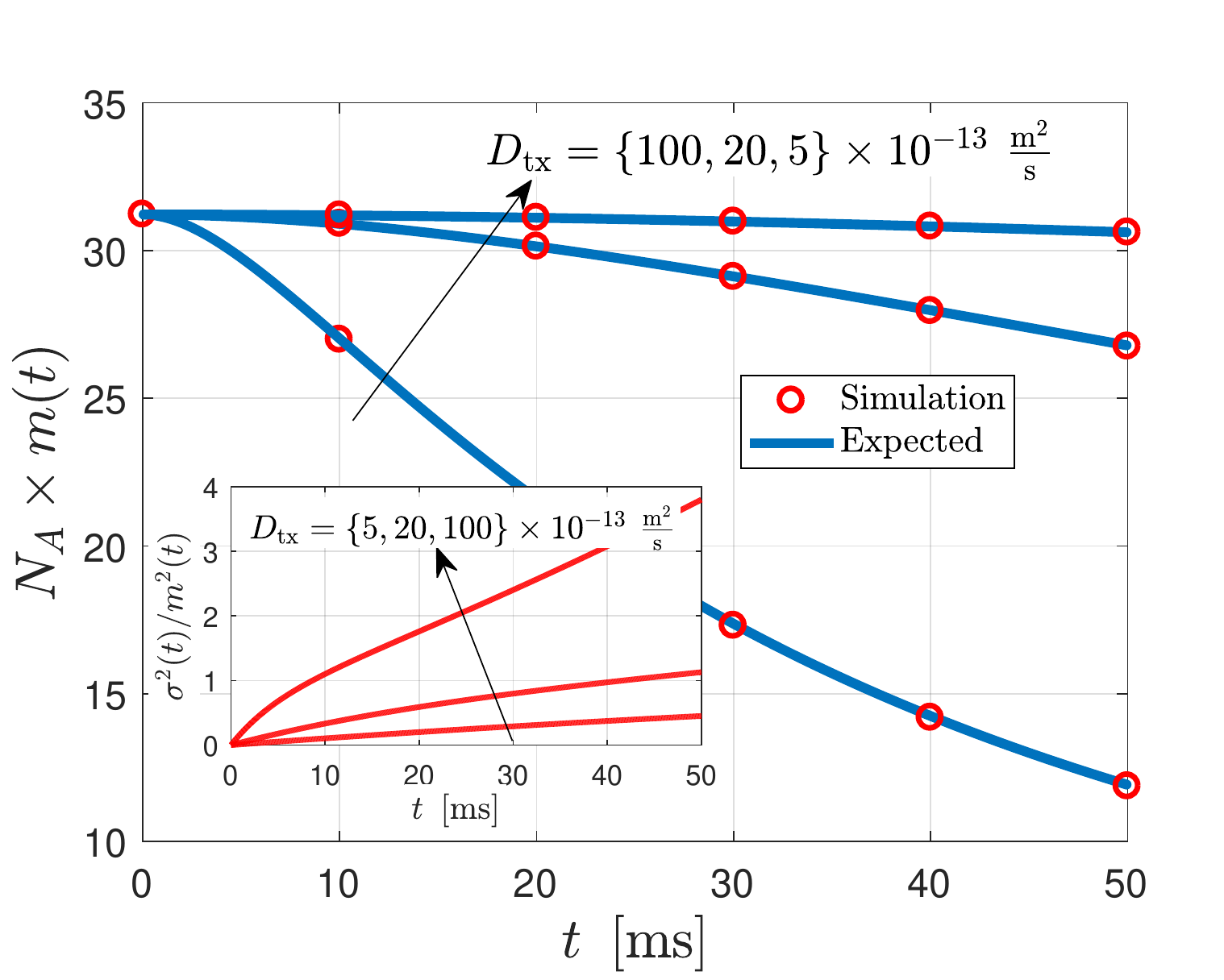}
	\caption{Expected received signal, $N_A m(t)$, as a function of time $t$ for different values of $\diffd{\TX}=\{5, 20, 100\}\times 10^{-13}$ $\frac{\text{m}^2}{\text{s}}$ in the absence of flow.}
	\label{Fig.Analysis2} 
\end{figure} 
In Fig.~\ref{Fig.Analysis2} and its inset, we investigate the impact of time $t$ on the mean and the normalized variance of the received signal in the absence of flow, i.e., $N_Am(t)$ and $\sigma^2(t) / m^2(t)$, respectively, for $\diffd{\TX}=\{5, 20, 100\}\times 10^{-13}$ $\frac{\text{m}^2}{\text{s}}$. Fig.~\ref{Fig.Analysis2} shows that as time $t$ increases, $N_A m(t)$ decreases. This is due to the fact that as $t$ increases, on average $r(t)$ increases as transmitter and receiver diffuse away and, consequently, $m(t)$ decreases. The decrease is faster for larger values of $\diffd{\TX}$, since for larger $\diffd{\TX}$, the transmitter diffuses away faster. The normalized variance of the received signal is shown in the inset of Fig.~\ref{Fig.Analysis2}. We observe that for all values of $\diffd{\TX}$, the normalized variance of the received signal is an increasing function of time. This is because as time increases, due to the Brownian motion of transmitter and receiver, the variance of their movements increases, which leads to an increase in the normalized variance of the received signal. As expected, this increase is faster for larger values of $\diffd{\TX}$, since the displacement variance of the transmitter, $2\diffd{\TX}t$, is larger.

\begin{figure}[!t]	
	\centering
	\includegraphics[scale=0.55]{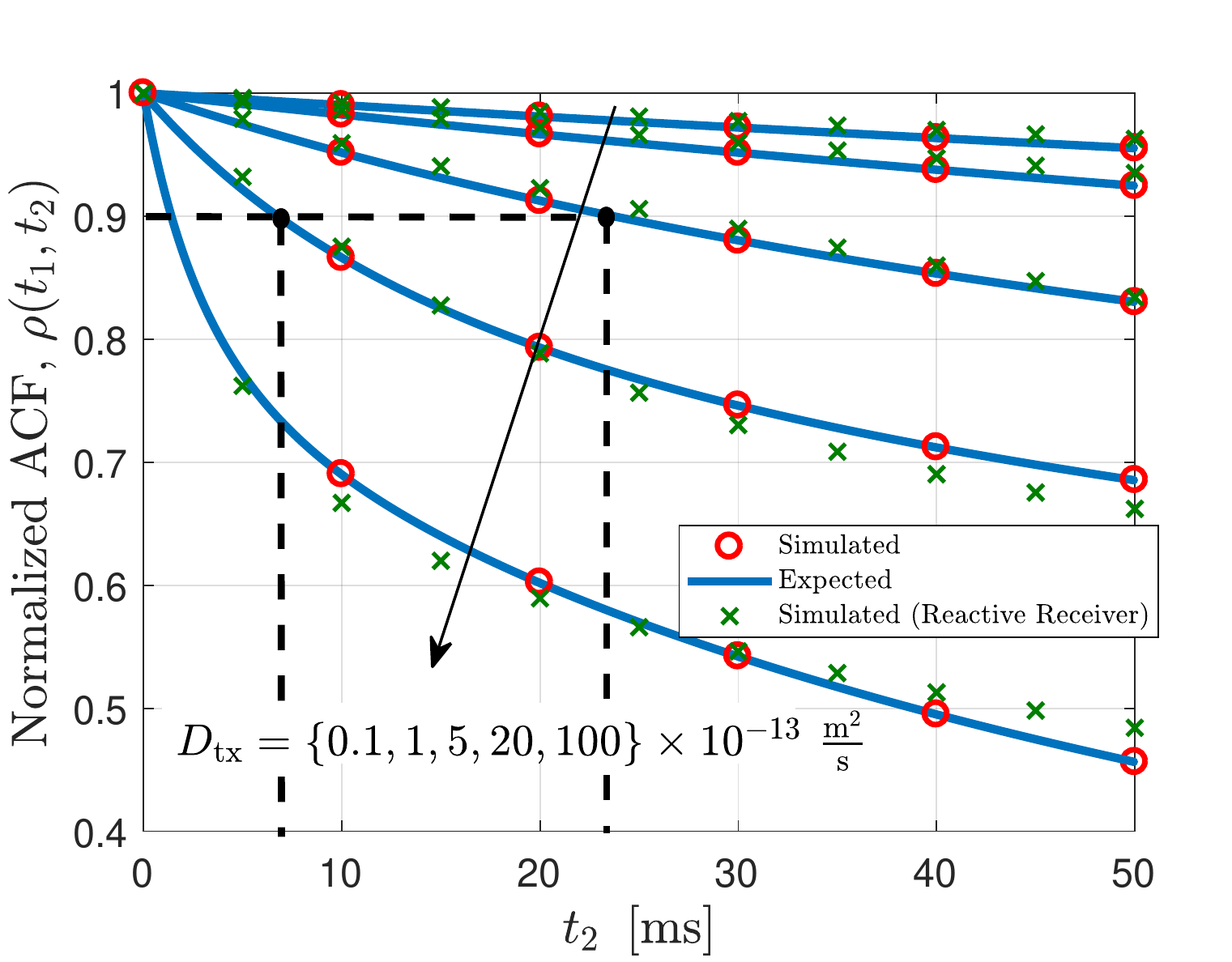}
	\caption{Normalized ACF, $\rho(t_1 = 0,t_2)$, as a function of $t_2$, for different values of $\diffd{\TX}=\{0.1, 1, 5, 20, 100\}\times 10^{-13}$ $\frac{\text{m}^2}{\text{s}}$ in the absence of flow.}
	\label{Fig.Analysis1}
\end{figure}
In Fig.~\ref{Fig.Analysis1}, the normalized ACF, $\rho(t_1,t_2)$, is evaluated as a function of $t_2$ in the absence of flow for a fixed value of $t_1 = 0$ and transmitter diffusion coefficients $\diffd{\TX}=\{0.1, 1, 5, 20, 100\}\times 10^{-13}$ $\frac{\text{m}^2}{\text{s}}$. We observe that for all considered values of $\diffd{\TX}$, $\rho(t_1,t_2)$ decreases with increasing $t_2$. This is due to the fact that by increasing $t_2$, on average $r(t)$ increases, and the CIR becomes more decorrelated from the CIR at time $t_1 = 0$. Furthermore, as expected, for larger values of $\diffd{\TX}$, $\rho(t_1,t_2)$ decreases faster, as for larger values of $\diffd{\TX}$, the transmitter diffuses away faster. For $\eta = 0.9$, the coherence time, $T^{\text{c}}$, for $\diffd{\TX} = 20 \times 10^{-13}$ $\frac{\text{m}^2}{\text{s}}$ and $\diffd{\TX} = 5 \times 10^{-13}$ $\frac{\text{m}^2}{\text{s}}$ is $7$ ms and $23$ ms, respectively. The coherence time is a measure for how frequently CSI acquisition has to be performed. In Fig.~\ref{Fig.Analysis1}, we also show the normalized ACF for the reactive receiver model developed in \cite{ArmanJ2} (cross markers). In particular, we assume that the surface of the reactive receiver is covered by $4000$ reciprocal receptors, each with radius $13.9$ nm. The binding and unbinding reaction rate constants of the signaling molecules to the receptor protein molecules are $1.25 \times 10^{-14}$ $\text{molecule}^{-1} \text{m}^3 \text{s}^{-1}$ and $2\times 10^{4}$ $\text{s}^{-1}$, respectively. Furthermore, for the reactive receiver scenario, we assume that the signaling molecules can degrade in the channel via a first-order degradation reaction with a reaction rate constant of $2\times 10^{4}$ $\text{s}^{-1}$. Fig.~\ref{Fig.Analysis1} shows that the analytical expression derived for the normalized ACF for the passive receiver constitutes a good approximation for the simulated normalized ACF for the reactive receiver. 

\begin{figure}[!t]
	\centering
	\includegraphics[scale=0.55]{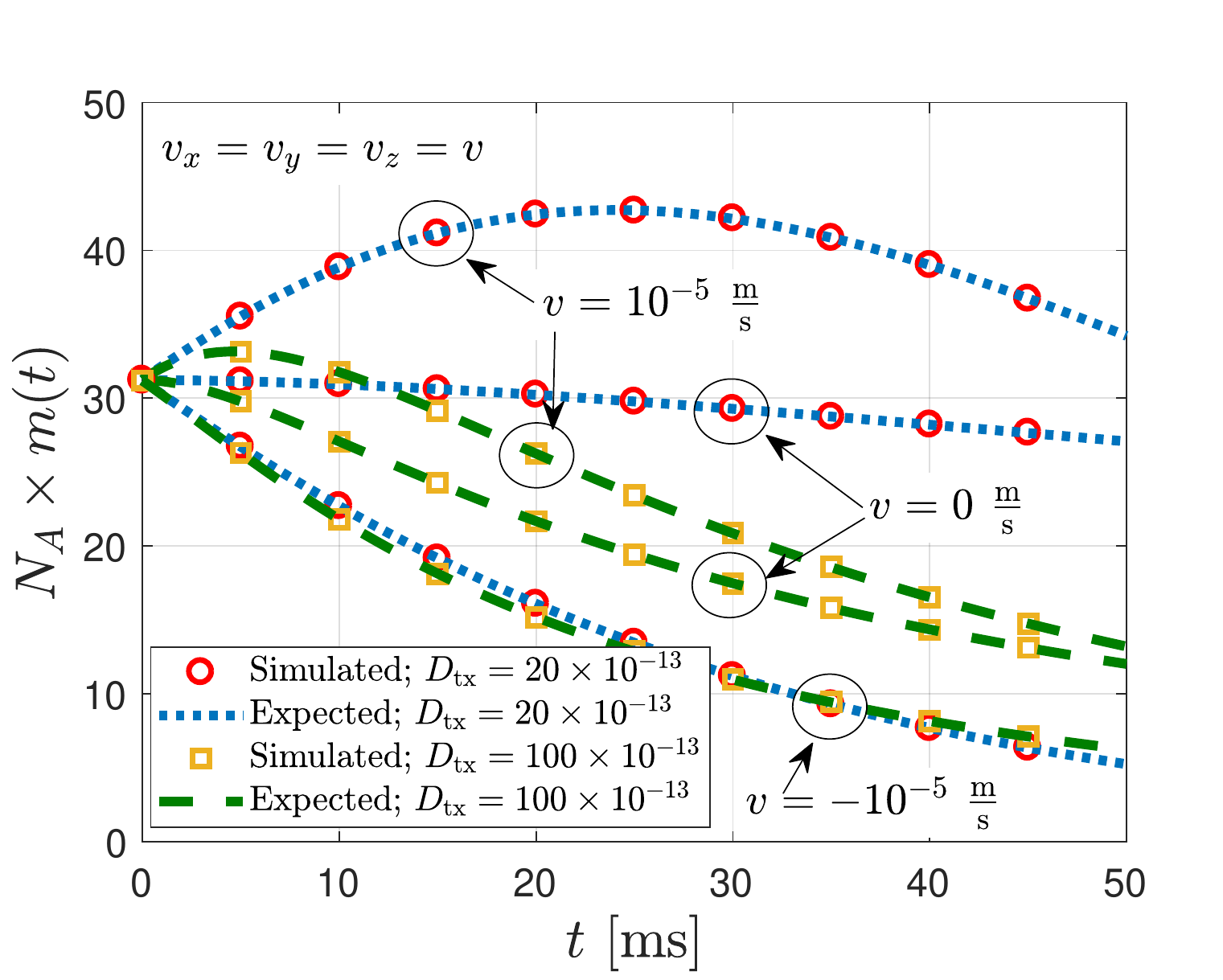}
	\caption{Expected received signal, $N_A m(t)$, as a function of time $t$ for a fixed receiver.}
	\label{Fig.Analysis3} 
\end{figure}
\begin{figure}[!t]
	\centering
	\includegraphics[scale=0.55]{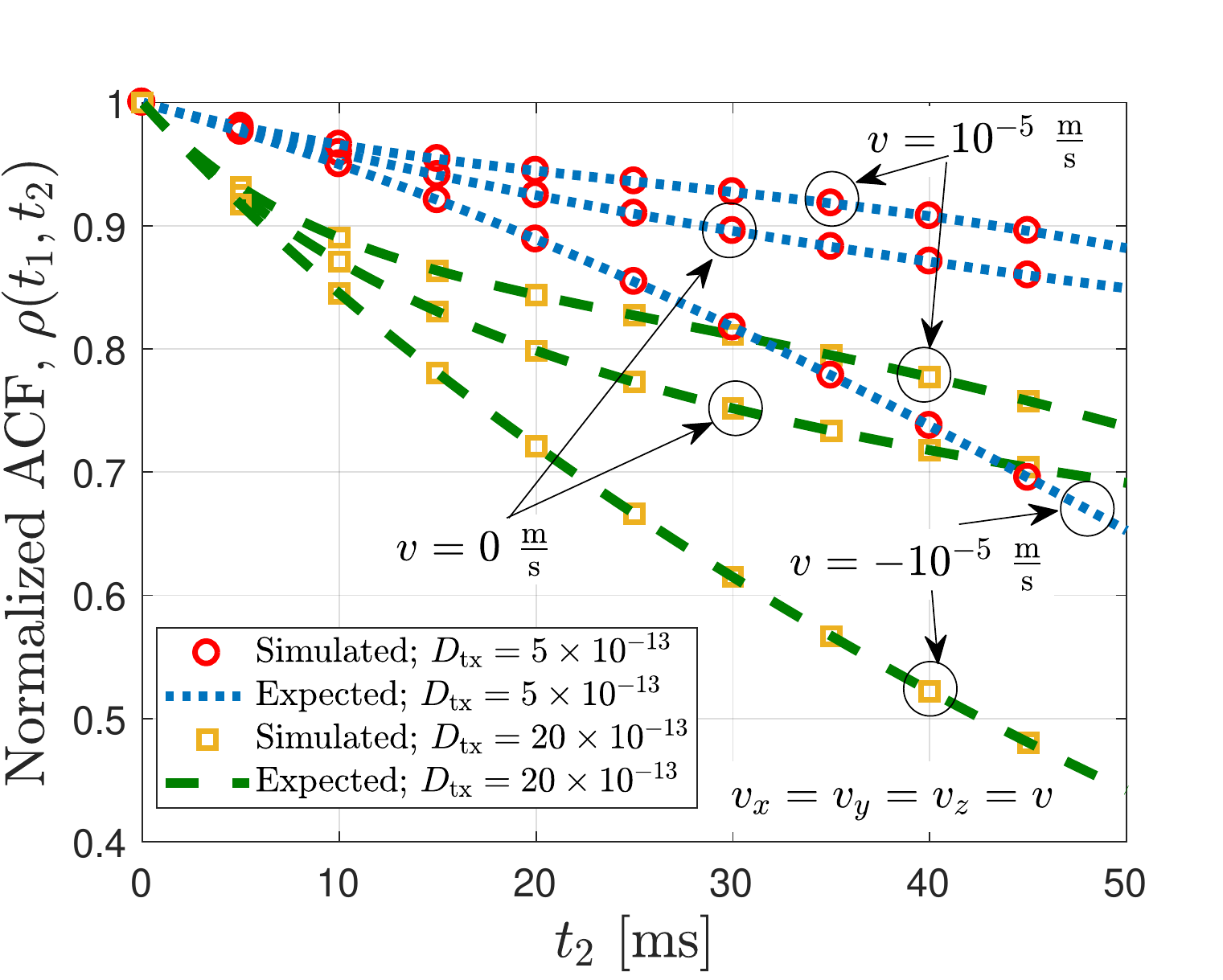}
	\caption{Normalized ACF, $\rho(t_1 = 0,t_2)$, as a function of $t_2$, for $\tau = \tau_s$, $\diffd{\TX}  =  \{ 5, 20\}\times 10^{-13}$ $\frac{\text{m}^2}{\text{s}}$, and $v \hspace{-1 mm} = \hspace{-1 mm} \{0,10^{-5},-10^{-5}\}$ $\frac{\mathrm{m}}{\mathrm{s}}$ for a fixed receiver.}
	\label{Fig.Analysis4}
\end{figure}

In Figs.~\ref{Fig.Analysis3} and \ref{Fig.Analysis4}, we investigate the impact of flow on $N_Am(t)$ and the normalized ACF. In Fig.~\ref{Fig.Analysis3}, $N_Am(t)$ is depicted as a function of time $t$ for system parameters $\diffd{\TX} = \{20,100 \} \times 10^{-13}$ $\frac{\mathrm{m}^2}{\mathrm{s}}$ and $v = \{0, 10^{-5}, -10^{-5} \}$ $\frac{\mathrm{m}}{\mathrm{s}}$, where we assumed $v_x = v_y = v_z = v$ and a fixed receiver. Fig.~\ref{Fig.Analysis3} shows that for positive $v$, $N_Am(t)$ first increases, as a positive flow carries the transmitter towards the receiver, and then decreases, since the transmitter eventually passes the receiver. Moreover, the increase of $N_Am(t)$ is larger for smaller values of $\diffd{\TX}$. This is because, when $\diffd{\TX}$ is small, flow is the dominant transport mechanism. However, when $\diffd{\TX}$ is large, diffusion becomes the dominant transport mechanism and, as discussed before, on average $r(t)$ increases, which reduces $N_Am(t)$. For the case when the flow is negative, $N_Am(t)$ decreases quickly. This behaviour is expected, as for $v < 0$, the flow carries the transmitter away from the receiver.                 

In Fig.~\ref{Fig.Analysis4}, the normalized ACF, $\rho(t_1,t_2)$, is evaluated as a function of $t_2$ for a fixed value of $t_1 = 0$, a fixed receiver, and system parameters $\diffd{\TX} = \{5,20\} \times 10^{-13}$ $\frac{\mathrm{m}^2}{\mathrm{s}}$ and $v = \{0, 10^{-5}, -10^{-5} \}$ $\frac{\mathrm{m}}{\mathrm{s}}$, where $v_x = v_y = v_z = v$. We observe that, for both considered values of $\diffd{\TX}$, if $v = 10^{-5}$ ($v = -10^{-5}$) $\frac{\mathrm{m}}{\mathrm{s}}$, $\rho(0,t_2)$ is larger (smaller) than the corresponding value when $v = 0$. This has the following reason. On the one hand, the variance of the movements of the transmitter in each Cartesian coordinate, $\sigma_{\TX}^2 = 2\diffd{\TX}t_2$, is an increasing function of time $t_2$ and \emph{independent} of $v$. On the other hand, for $v = 10^{-5}$ ($v = -10^{-5}$) $\frac{\mathrm{m}}{\mathrm{s}}$, on average the transmitter is closer to (farther from) the receiver than for $v = 0$ $\frac{\mathrm{m}}{\mathrm{s}}$. Thus, at any given time $t_2$, $\sigma_{\TX}^2$ leads to relatively smaller (larger) variations of $h(t_2,\tau_s)$ for the case when $v = 10^{-5}$ ($v = -10^{-5}$) $\frac{\mathrm{m}}{\mathrm{s}}$ compared with the case when $v = 0$ $\frac{\mathrm{m}}{\mathrm{s}}$. This leads to a larger (smaller) value of $\rho(0,t_2)$ for $v = 10^{-5}$ ($v = -10^{-5}$) $\frac{\mathrm{m}}{\mathrm{s}}$ than for $v = 0$ $\frac{\mathrm{m}}{\mathrm{s}}$.  
        
We note the excellent match between simulation and analytical results in Figs.~3-6. 

\subsection{CDF and PDF of CIR}
\label{Sec.SimCDFPDF}
\begin{figure}[!t]
	\centering
	\includegraphics[scale=0.55]{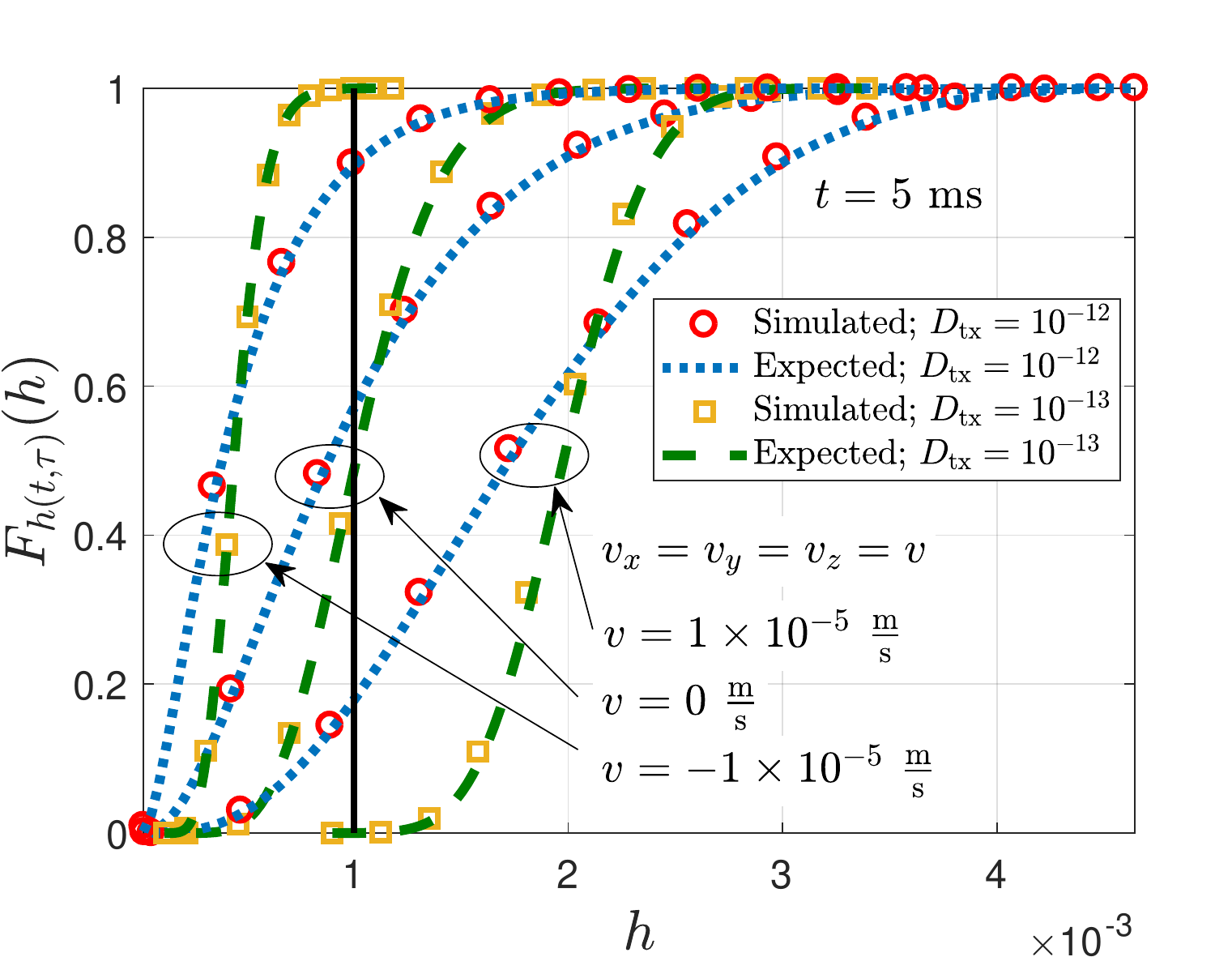}
	\caption{CDF of the CIR, $\textit{F}_{h(t,\tau)}(h)$, at $t=5$ ms.}
	\label{Fig.CDFAnalysis}
\end{figure}
In Fig.~\ref{Fig.CDFAnalysis}, the CDF of the impulse response of a time-variant MC channel, $\textit{{\large F}}_{h(t,\tau)}(h)$, is shown for system parameters $\diffd{\text{tx}} = \{10^{-12}, 10^{-13} \}$ $\frac{\text{m}^2}{\text{s}}$, $v_x = v_y = v_z = v = \{0, 10^{-5}, -10^{-5}\}$ $\frac{\mathrm{m}}{\mathrm{s}}$, a fixed receiver, and time $t = 5$ ms, i.e., $t = 10T$. We observe that for all considered values of $v$, increasing $\diffd{\text{tx}}$ makes the CDF wider, as for larger values of $\diffd{\text{tx}}$ the variance of the movements of the transmitter and, as a result, the variance of $\vec{r}(t)$ increase, which leads to an increase in the variance of $h(t,\tau)$. Furthermore, Fig.~\ref{Fig.CDFAnalysis} shows that for a given $\diffd{\text{tx}}$ and a fixed receiver, a positive and a negative flow shift the CDF of the CIR to the right and the left, respectively, compared to the case without flow. This is because, e.g. in the presence of a positive flow, the transmitter is pushed towards the receiver and hence, $\vec{r}(t)$ decreases. As a result, larger values of $h$ are more likely to occur. Furthermore, the solid black line in Fig.~\ref{Fig.CDFAnalysis} denotes $h = h_{\text{min}} = 10^{-3}$, which corresponds to an average error probability of approximately $10^{-3}$. Fig.~\ref{Fig.CDFAnalysis} reveals that after $5$ ms, i.e., after transmission of $10$ bits, the outage probability is higher for $v = 0$ $\frac{\mathrm{m}}{\mathrm{s}}$ and $v = -10^{-5}$ $\frac{\mathrm{m}}{\mathrm{s}}$ compared to $v = 10^{-5}$ $\frac{\mathrm{m}}{\mathrm{s}}$, as for $v = \{0, -10^{-5}\}$ $\frac{\mathrm{m}}{\mathrm{s}}$, transmitter and receiver are on average further apart after $5$ ms. We note again the excellent matched between simulation and analytical results. 

\begin{figure}[!t]	
	\centering
	\includegraphics[scale=0.55]{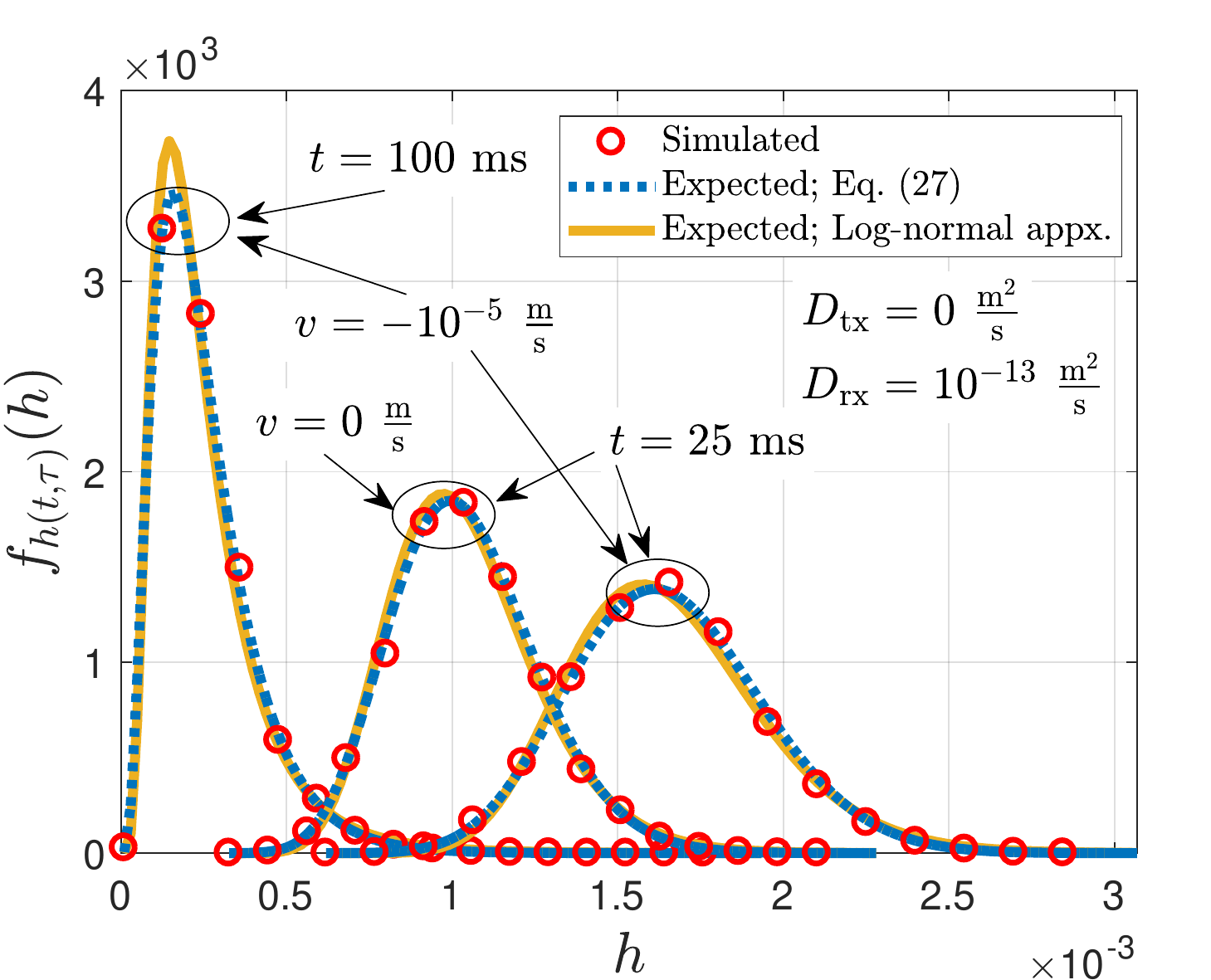}
	\caption{PDF of the CIR, $\textit{f}_{h(t,\tau)}(h)$, for $v = \{0, -10^{-5}\}$ $\frac{\mathrm{m}}{\mathrm{s}}$ and $t = \{25, 100\}$ ms.}
	\label{Fig.PDFAnalysis1}
\end{figure}
In Fig.~\ref{Fig.PDFAnalysis1}, the PDF of the time-variant CIR, $\textit{{\large f}}_{h(t,\tau)}(h)$, is evaluated for system parameters $t = \{25, 100\}$ ms, $v_x = v_y = v_z = v = \{0,-10^{-5}\}$ $\frac{\mathrm{m}}{\mathrm{s}}$, $\diffd{\RX} = 10^{-13}$ $\frac{\mathrm{m}^2}{\mathrm{s}}$, and a \emph{fixed transmitter}. For the case of a fixed transmitter, in the presence of a negative flow, e.g., $v = -10^{-5}$ $\frac{\mathrm{m}}{\mathrm{s}}$, the receiver first moves towards the transmitter before passing it. Thus, as shown in Fig.~\ref{Fig.PDFAnalysis1}, first, for $t = 25$ ms, $\textit{{\large f}}_{h(t,\tau)}(h)$ is shifted to the right, and later for $t = 100$ ms, when the receive is far away from the transmitter, $\textit{{\large f}}_{h(t,\tau)}(h)$ is shifted to the left. We note the excellent agreement of the derived expression for $\textit{{\large f}}_{h(t,\tau)}(h)$, i.e., \eqref{Eq.PDFofCIR} with the simulation results. We also observe that the Log-normal distribution provides a good approximation for the PDF of the CIR in Fig.~\ref{Fig.PDFAnalysis1}, as for all three considered cases, the necessary condition \eqref{Eq.NecCondForLogNormalApp} is satisfied.       

\begin{figure}[!t]
	\centering
	\includegraphics[scale=0.55]{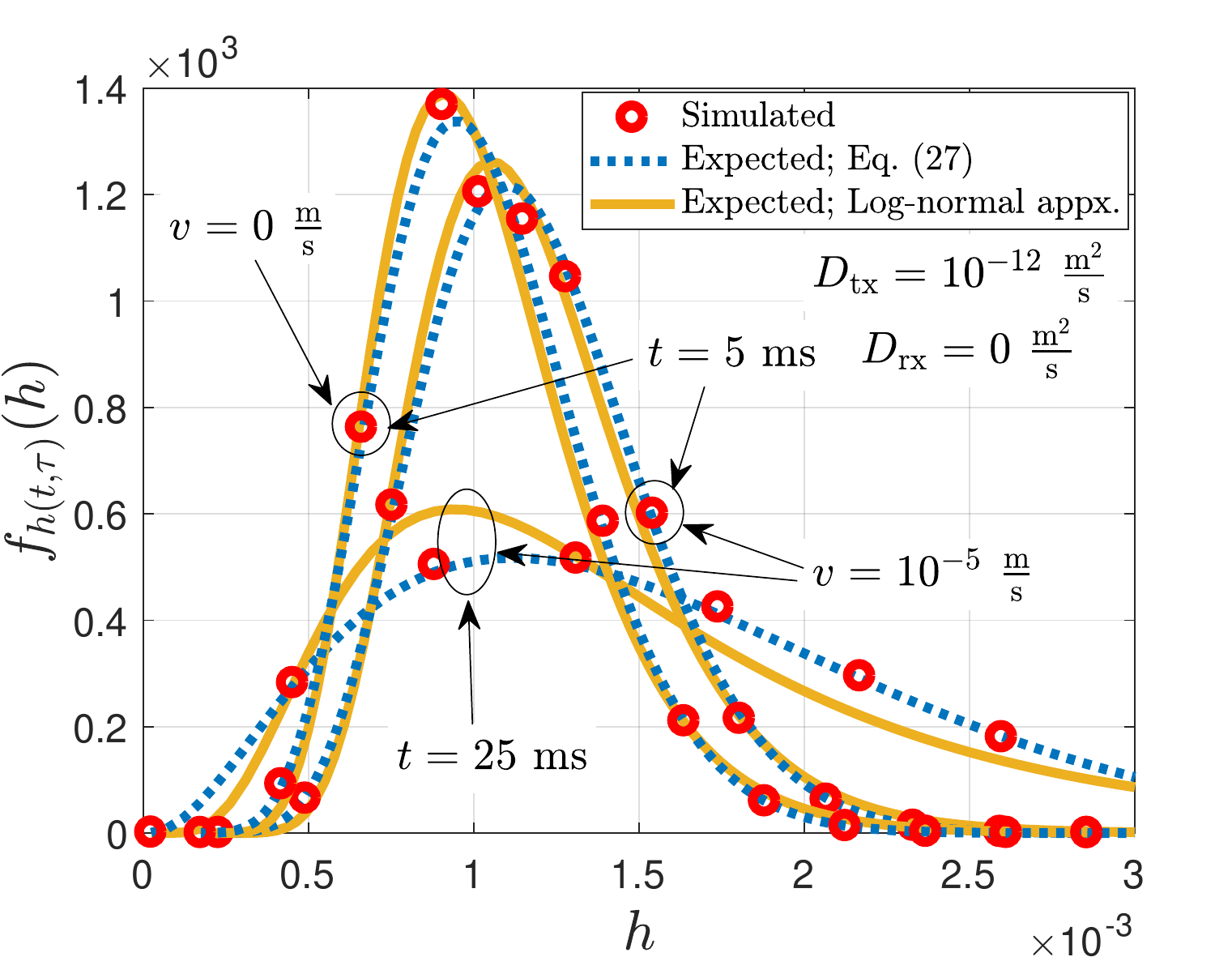}
	\caption{PDF of the CIR, $\textit{f}_{h(t,\tau)}(h)$, for $v = \{0, 10^{-5}\}$ $\frac{\mathrm{m}}{\mathrm{s}}$ and $t = \{5, 25\}$ ms.}
	\label{Fig.PDFAnalysis2}
\end{figure}
In Fig.~\ref{Fig.PDFAnalysis2}, $\textit{{\large f}}_{h(t,\tau)}(h)$ is depicted for system parameters $t = \{5, 25\}$ ms, $v_x = v_y = v_z = v = \{0,-10^{-5}\}$ $\frac{\mathrm{m}}{\mathrm{s}}$, $\diffd{\TX} = 10^{-12}$ $\frac{\mathrm{m}^2}{\mathrm{s}}$, and a \emph{fixed receiver}. In Fig.~\ref{Fig.PDFAnalysis2}, since the receiver is fixed, in the presence of a positive flow $v = 10^{-5}$ $\frac{\mathrm{m}}{\mathrm{s}}$, the transmitter moves towards the receiver, and hence, $\textit{{\large f}}_{h(t,\tau)}(h)$ shifts to the right compared to the case without flow, see e.g. for time $t = 5$ ms. As time increases, the transmitter passes the receiver and $r(t)$ starts to increase, and hence, $\textit{{\large f}}_{h(t,\tau)}(h)$ starts to shift to the left. However, in Fig.~\ref{Fig.PDFAnalysis2}, the effective diffusion coefficient $D_2 = \diffd{\TX} = 10^{-12}$ $\frac{\mathrm{m}^2}{\mathrm{s}}$ is greater than the effective diffusion coefficient $D_2 = \diffd{\RX} = 10^{-13}$ $\frac{\mathrm{m}^2}{\mathrm{s}}$ in Fig.~\ref{Fig.PDFAnalysis1}. As a result, the Log-normal distribution approximation starts to deviate from the actual PDF sooner, i.e., at $t = 25$ ms. This is because for larger values of $D_2$, condition \eqref{Eq.NecCondForLogNormalApp} is violated for smaller $t$.         

\begin{figure}[!t]	
	\centering
	\includegraphics[scale=0.55]{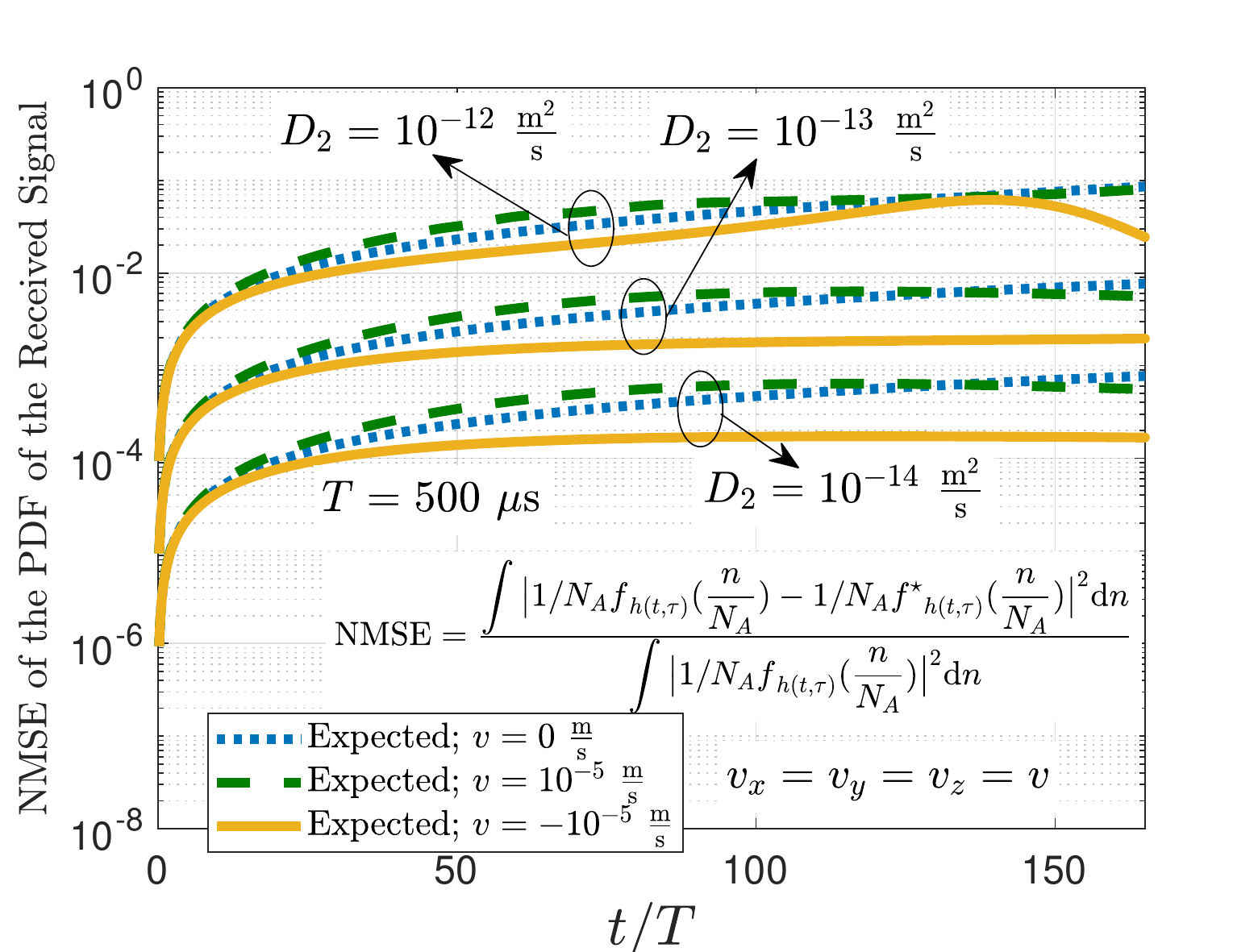}
	\caption{NMSE of the PDF of the received signal as a function of dimensionless time, $t/T$.}
	\label{Fig.NMSE_PDF_APP}
\end{figure}
In order to evaluate the accuracy of the proposed approximate PDF, $\textit{{\large f}}_{h(t,\tau)}^{\,\,\star}(h) $, in Fig.~\ref{Fig.NMSE_PDF_APP}, the NMSE of the PDF of the received signal, defined as $\mathrm{NMSE} = \big( \int |1/N_A \textit{{\large f}}_{h(t,\tau)}(\frac{n}{N_A}) - 1/N_A \textit{{\large f}}_{h(t,\tau)}^{\,\,\star}(\frac{n}{N_A})|^2 \mathrm{d}n \big/ \int |1/N_A \textit{{\large f}}_{h(t,\tau)}(\frac{n}{N_A})|^2 \mathrm{d}n \big)$, is evaluated as a function of normalized time, $t/T$, for system parameters $D_2 = \{10^{-12}, 10^{-13}, 10^{-14}\}$ $\frac{\mathrm{m}^2}{\mathrm{s}}$ and $v = \{0,10^{-5}, -10^{-5}\}$ $\frac{\mathrm{m}}{\mathrm{s}}$ for a fixed receiver. Fig.~\ref{Fig.NMSE_PDF_APP} shows that, for a given time $t$, the NMSE grows with the effective diffusion coefficient of transmitter and receiver, $D_2$. This is because condition \eqref{Eq.NecCondForLogNormalApp} is inversely proportional to $D_2$. In other words, for smaller values of $D_2$, the maximum time, $t_{\mathrm{max}}$, that satisfies \eqref{Eq.NecCondForLogNormalApp} is larger than for larger values of $D_2$. For example, in Fig.~\ref{Fig.NMSE_PDF_APP}, for $D_2 = 10^{-13}$ $\frac{\mathrm{m}^2}{\mathrm{s}}$ and $v = 0$ $\frac{\mathrm{m}}{\mathrm{s}}$, $t_{\mathrm{max}} \simeq 67T = 33.5$ ms, while for $D_2 = 10^{-12}$ $\frac{\mathrm{m}^2}{\mathrm{s}}$ and $v = 0$ $\frac{\mathrm{m}}{\mathrm{s}}$, $t_{\mathrm{max}} \simeq 7T = 3.5$ ms. Furthermore, we can observe that for the considered values of $D_2$, when $v < 0$, NMSE is smaller compared to the case when $v \geq 0$. This is because for $v < 0$, $r^{\text{eq}}(t)$ in \eqref{Eq.NecCondForLogNormalApp} increases more quickly over time, which yields smaller values of NMSE.            
    
\subsection{Error Rate Analysis}
\begin{figure}[!t]
	\centering
	\includegraphics[scale=0.55]{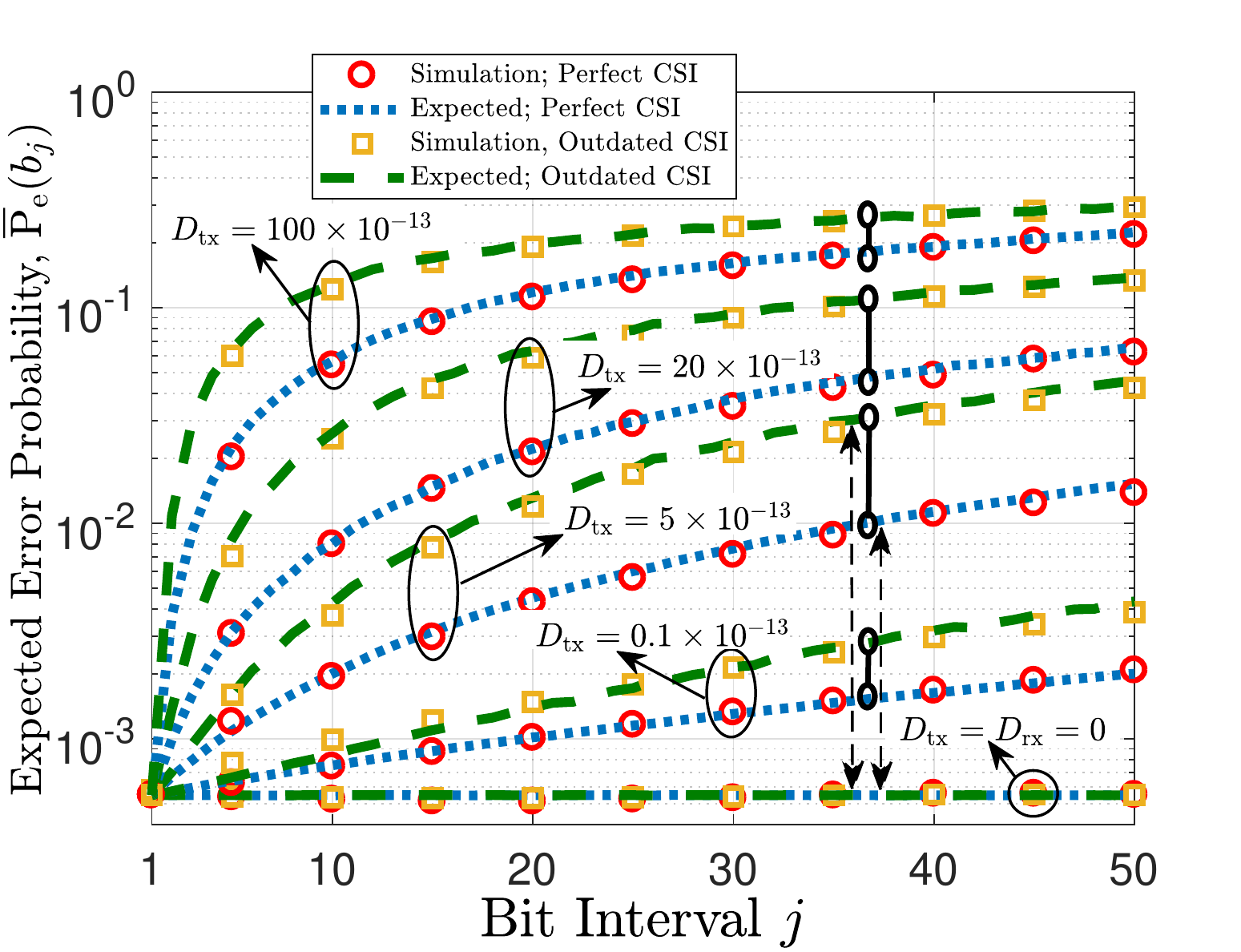}
	\caption{Expected error probability, $\mathrm{\overline{P}_e}(b_j)$, as a function of bit interval $j$.}
	\label{Fig.BERAnalysis1}
\end{figure}
In Fig.~\ref{Fig.BERAnalysis1}, the expected error probability, $\mathrm{\overline{P}_e}(b_j)$, is shown as a function of bit interval $j$ in the absence of flow for system parameters $\diffd{\TX}=\{0.1, 5, 20, 100\}\times 10^{-13}$ $\frac{\text{m}^2}{\text{s}}$ as well as for the conventional case of fixed transmitter and fixed receiver, i.e., $\diffd{\TX} = \diffd{\RX} = 0$. As expected, when transmitter and receiver are fixed, the performances of the detectors with perfect and outdated CSI are identical, as the channel does not change over time. On the other hand, when $\diffd{\TX} > 0$, the performances of both detectors deteriorate over time. This is due to the fact that as time increases, i) $\sigma^2(t)$ increases and ii) $m(t)$ decreases. Furthermore, the gap between the BERs of the detector with perfect CSI and the detector with outdated CSI increases over time since the impulse response of the channel decorrelates (see Fig.~\ref{Fig.Analysis1}), and, as a result, the CSI becomes outdated. Moreover, the CSI becomes outdated faster for larger values of $\diffd{\TX}$. Hence, for a given time (bit interval), the absolute value of the performance gap between both cases, highlighted by solid black lines in Fig.~\ref{Fig.BERAnalysis1}, increases. For instance, for $j = 37$, the absolute values of the performance gaps between the detectors with perfect and outdated CSI for $\diffd{\TX} = \{ 0.1, 5, 20, 100\}\times 10^{-13}$ $\frac{\text{m}^2}{\text{s}}$ are $\{0.0013, 0.0212, 0.0624, 0.08\}$, respectively. 

\begin{figure}[!t]	
	\centering
	\includegraphics[scale=0.55]{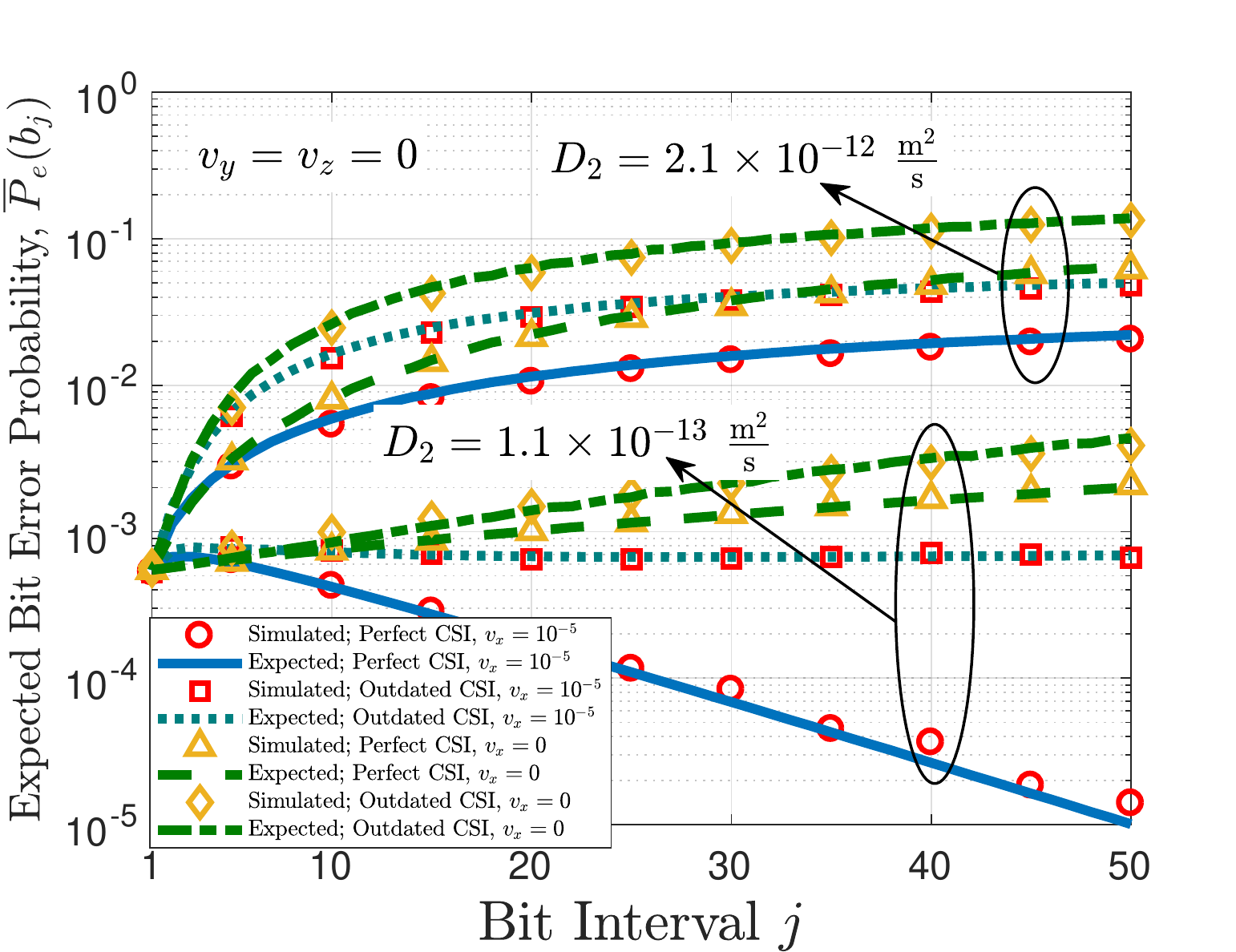}
	\caption{$\mathrm{\overline{P}_e}(b_j)$ as a function of bit interval $j$ for $v_x = \{0,10^{-5}\}$ $\frac{\mathrm{m}}{\mathrm{s}}$ and $D_2 = \{1.1, 21 \} \times 10^{-13}$ $\frac{\mathrm{m}^2}{\mathrm{s}}$.}  
	\label{Fig.BERAnalysis2}
\end{figure}
\begin{figure}[!t]
	\centering
	\includegraphics[scale=0.55]{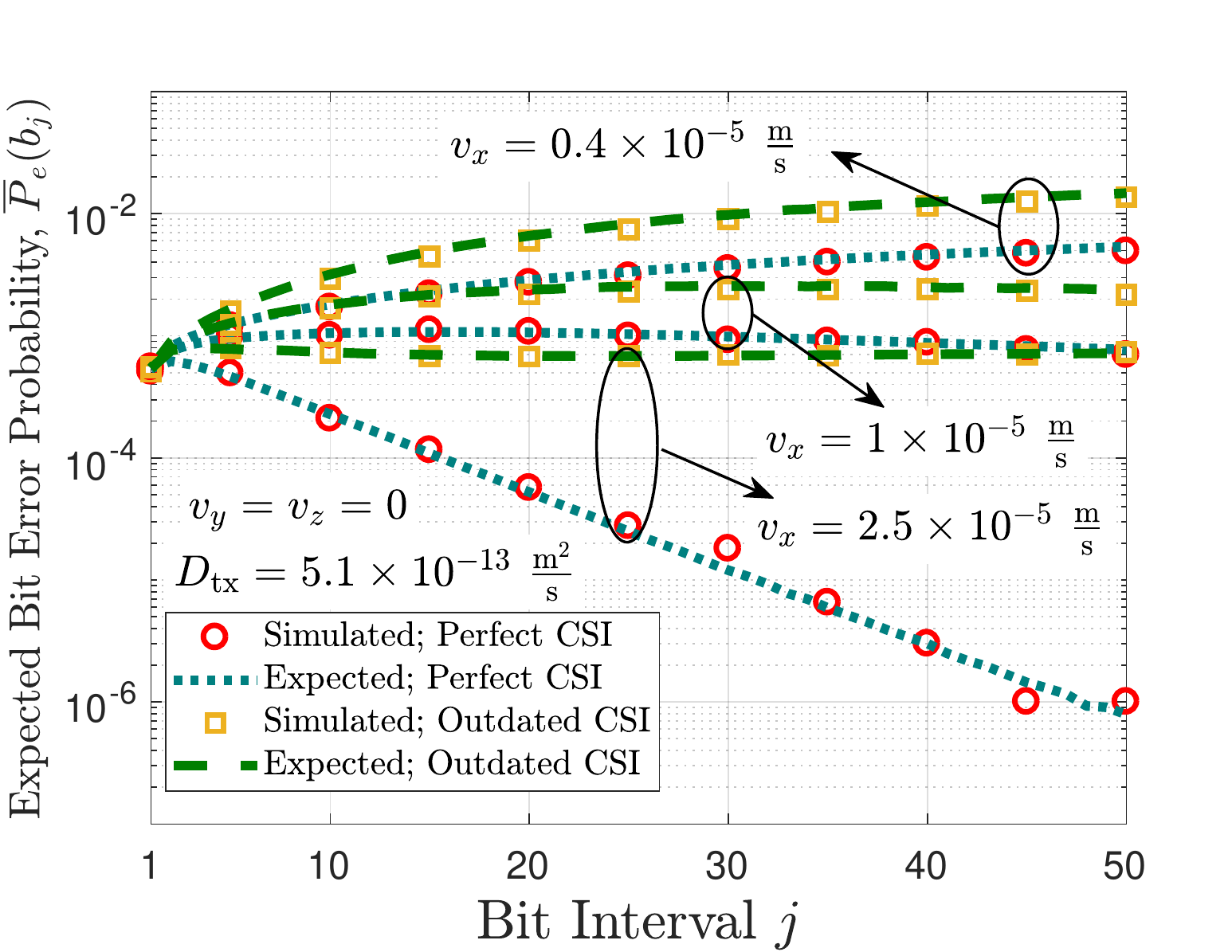}
	\caption{Expected error probability, $\mathrm{\overline{P}_e}(b_j)$, as a function of bit interval $j$.}
	\label{Fig.BERAnalysis3}
\end{figure}

In Figs.~\ref{Fig.BERAnalysis2} and \ref{Fig.BERAnalysis3}, we study the impact of flow on the performance of mobile MC systems. The dimensionless Peclet number, $\mathrm{Pe_L}$, which quantifies the relative importance of convection and diffusion for molecule transport, is considered as a metric for characterization of the channel. Here, $\mathrm{Pe_L}$ is defined as $\frac{L_{\mathrm{ref}}|\vec{v}|}{D_2}$, where $L_{\mathrm{ref}}$ is the characteristic length scale, which we choose as $L_{\mathrm{ref}} = r_0/2$. This corresponds to a 50\% reduction of the initial distance between transmitter and receiver. Values of $\mathrm{Pe_L} \gg 1$ (in practice $\mathrm{Pe_L} > 10$) correspond to a regime where the impact of flow is dominant compared to that of diffusion, whereas values of $\mathrm{Pe_L} \ll 1$ (in practice $\mathrm{Pe_L} < 0.1$) correspond to a diffusion dominated regime. Furthermore, values of $0.1 < \mathrm{Pe_L} < 10$ correspond to a regime, which we refer to as ``intermediate regime'' and where neither diffusion nor flow are dominant, see e.g. \cite{Benoit}.
   
In Fig.~\ref{Fig.BERAnalysis2}, $\mathrm{\overline{P}_e}(b_j)$ is evaluated as a function of bit interval $j$ for two cases, with and without flow. For the case of without flow, where $\mathrm{Pe_L}=0$, we consider a mobile transmitter and receiver, and adopt $\diffd{\TX} = \{0.1, 20\} \times 10^{-13}$ $\frac{\mathrm{m}^2}{\mathrm{s}}$, $\diffd{\RX} = 10^{-13}$ $\frac{\mathrm{m}^2}{\mathrm{s}}$, and $v_y = v_z = v_x = 0$ $\frac{\mathrm{m}}{\mathrm{s}}$. For the second case, we consider a mobile transmitter and a fixed receiver in the presence of flow, and assume $\diffd{\RX} = 0$ $\frac{\mathrm{m}^2}{\mathrm{s}}$, $v_y = v_z = 0$ $\frac{\mathrm{m}}{\mathrm{s}}$, and $v_x = 10^{-5}$ $\frac{\mathrm{m}}{\mathrm{s}}$. Furthermore, in order to have a fair comparison between both cases with respect to $\diffd{2}$, for the second case, we adopt $\diffd{\TX} = \{1.1, 21\}\times 10^{-13}$ $\frac{\mathrm{m}^2}{\mathrm{s}}$ such that the same value of $D_2$ results for both cases. We note that, for the second case, $D_{\mathrm{tx}} = 21 \times 10^{-13}$ $\frac{\mathrm{m}^2}{\mathrm{s}}$ corresponds to $ 0.1 < \mathrm{Pe_L} = 2.38 < 10$, whereas $D_{\mathrm{tx}} = 1.1 \times 10^{-13}$ $\frac{\mathrm{m}^2}{\mathrm{s}}$ corresponds to $\mathrm{Pe_L} = 45.45 > 10$. First of all, Fig.~\ref{Fig.BERAnalysis2} shows that for both considered values of $D_2$, by increasing $v_x$ the performance of both detectors improves. This is because in the presence of positive flow $v_x = 10^{-5}$ $\frac{\mathrm{m}}{\mathrm{s}}$, $r(t)$ increases on average later in time compared to the case without flow, since the transmitter is moved towards the receiver, and, as a result, the performance of both detectors improves. Furthermore, we can see that the performance gap between the two detectors becomes larger as $\diffd{\TX}$ decreases.  This is because for the detector with outdated CSI, the channel does not only decorrelate over time, but the mean of the channel also changes drastically for smaller values of $D_2$ in the presence of positive flow compared with the case without flow, and as a result, the performance gap between the two detectors is larger for smaller values of $D_2$.

In Fig.~\ref{Fig.BERAnalysis3}, the impact of flow on $\mathrm{\overline{P}_e}(b_j)$ is investigated for system parameters $\diffd{\TX} = 5.1 \times 10^{-13}$ $\frac{\mathrm{m}^2}{\mathrm{s}}$, $v_y = v_z = 0$ $\frac{\mathrm{m}}{\mathrm{s}}$, and $v_x = \{0.4, 1, 2.5 \} \times 10^{-5}$ $\frac{\mathrm{m}}{\mathrm{s}}$. Interestingly, Fig.~\ref{Fig.BERAnalysis3} reveals that when diffusion is slightly dominant over the flow in the intermediate regime ($v_x = 0.4 \times 10^{-5}$ $\frac{\mathrm{m}}{\mathrm{s}}$, $\mathrm{Pe_L = 3.9}$), the performance of both detection schemes gradually deteriorates over time, as on average $r(t)$ increases. However, for $\mathrm{Pe_L} \simeq 10$ ($v_x = 1 \times 10^{-5}$ $\frac{\mathrm{m}}{\mathrm{s}}$, $\mathrm{Pe_L} = 9.8$), diffusion and flow essentially cancel out each others' impact and  $r(t)$ remains on average approximately constant. Thus, the BERs of both detectors also remain approximately constant over time. However, in a flow dominated regime ($v_x = 2.5 \times 10^{-5}$ $\frac{\mathrm{m}}{\mathrm{s}}$, $\mathrm{Pe_L} = 24.5$), since $r(t)$ decreases on average over time for the duration of the considered bit intervals, the BER for perfect CSI decreases over time but the BER for outdated CSI still increases because of the inaccurate decision threshold.   

\section{Conclusions}
\label{Sec.Con} 
In this paper, we established a mathematical framework for the statistical characterization of the time-variant CIR of mobile MC channels. In particular, we derived closed-form expressions for the mean, the ACF, the CDF, and the PDF of the time-variant CIR. Furthermore, we approximated the PDF of the CIR by a Log-normal distribution, quantified the regime where this approximation is valid, and proposed a simple model for outdated CSI. Our analytical and simulation results reveal that the coherence time of the channel decreases when transmitter and/or receiver diffuse faster. Furthermore, our results show that outages are more likely to occur when flow causes the transmitter and receiver to drift apart and/or transmitter and receiver diffuse faster. The presented analysis also reveals that the accuracy of the Log-normal approximation of the PDF of the CIR decreases slower over time for smaller effective diffusion coefficients of transmitter and receiver. In addition, we have confirmed that both CIR decorrelation over time and flow influence the performance gap between detectors having perfect and outdated CSI. Overall, our results show that new modulation, detection, and estimation techniques have to be developed to enable reliable communication over time-variant mobile MC channels.

\appendices
\section{Proof of Theorem 2}
\label{App.1}
\newcounter{tempequationcounter}
\begin{figure*}[!t]
\normalsize
\setcounter{tempequationcounter}{\value{equation}}
\begin{IEEEeqnarray}{C} 
\label{Eq.Param}
\setcounter{equation}{47}
\vartheta = 2\left(\alpha + \beta(t_2 - t_1) + \beta(t_1)\right),\,\,\, \varepsilon = 2 \left( \alpha + \beta(t_2 - t_1) \right),\,\,\, \psi  = -2 \beta(t_2 - t_1) \nonumber \\
\mu_{x_1} = \frac{2 \left[ \frac{\varepsilon - \psi}{\varepsilon} (\alpha \vp{x}\tau) + \frac{\varepsilon + \psi}{\varepsilon} \left(\beta(t_2 - t_1)\vs{x}(t_2 - t_1) \right) + \beta(t_1) (x_0 - \vs{x}t_1) \right]}{\vartheta - \psi^2/\varepsilon}, \nonumber \\
\mu_{y_1} = \frac{2 \left[ \frac{\varepsilon - \psi}{\varepsilon} (\alpha \vp{y}\tau) + \frac{\varepsilon + \psi}{\varepsilon} \left(\beta(t_2 - t_1)\vs{y}(t_2 - t_1) \right) -\beta(t_1) \vs{y}t_1 \right]}{\vartheta - \psi^2/\varepsilon},  \nonumber \\
\mu_{z_1} = \frac{2 \left[ \frac{\varepsilon - \psi}{\varepsilon} (\alpha \vp{z}\tau) + \frac{\varepsilon + \psi}{\varepsilon} \left(\beta(t_2 - t_1)\vs{z}(t_2 - t_1) \right) -\beta(t_1) \vs{z}t_1 \right]}{\vartheta - \psi^2/\varepsilon},  \nonumber \\
\mu_{x_2} = \frac{2\alpha \vp{x}\tau -2\beta(t_2 - t_1)\vs{x}(t_2 - t_1) -\psi \mu_{x_1}}{\varepsilon},\,\,\, \mu_{y_2} = \frac{2\alpha \vp{y}\tau -2\beta(t_2 - t_1)\vs{y}(t_2 - t_1) -\psi \mu_{y_1}}{\varepsilon}, \nonumber \\
\mu_{z_2} = \frac{2\alpha \vp{z}\tau -2\beta(t_2 - t_1)\vs{z}(t_2 - t_1) -\psi \mu_{z_1}}{\varepsilon}.  
\end{IEEEeqnarray}
\setcounter{equation}{\value{tempequationcounter}}
\hrulefill
\end{figure*}
Given \eqref{Eq.jointPDF}, substituting $h(t_1,\tau)\big|_{\vec{r}(t) = \vec{r}_1}$ and $h(t_2,\tau)\big|_{\vec{r}(t) = \vec{r}_2}$ from \eqref{Eq.CIRTimVar} in \eqref{Eq.ACF_def}, we can write $\phi(t_1,t_2)$ as 
\begin{IEEEeqnarray}{rCl}
	\label{Eq.ACF_Proof1}
	\phi(t_1,t_2) & = & \varphi^2 \lambda(t_2 - t_1) \lambda(t_1) \hs \iint \limits_{\vec{r}_1,\,\vec{r}_2 \in \mathbb{R}^3} \hs e^{-\beta(t_2 - t_1) \left|\vec{r}_2 - (\vec{r}_1 - \vecs (t_2 - t_1))\right|^2} \nonumber \\
	&& \times\> e^{-\beta(t_1) \left|\vec{r}_1 - (\vec{r}_0 - \vecs t_1)\right|^2} e^{-\alpha \left|\vec{r}_1 - \vecp \tau \right|^2}  e^{-\alpha \left|\vec{r}_2 - \vecp \tau \right|^2}   \dif \vec{r}_2 \dif \vec{r}_1. \nonumber \\* 
\end{IEEEeqnarray} 
Expanding the integrands in \eqref{Eq.ACF_Proof1} leads to
\begin{IEEEeqnarray}{rCl}
	\label{Eq.ACF_Proof2} 
\phi(t_1,t_2) & = & \varphi^2 \lambda(t_2 - t_1) \lambda(t_1) \int_{-\infty}^{+\infty} \cdots \int_{-\infty}^{+\infty} e^{-\alpha \left(x_1 - \vp{x}\tau \right)^2} \nonumber \\ 
&& \times\> e^{-\alpha \left(x_2 - \vp{x}\tau \right)^2 -\beta(t_2 - t_1)\left(x_2 - x_1 + \vs{x}(t_2 - t_1)\right)^2} \nonumber \\
&& \times\> e^{-\beta(t_1) \left( x_1 - x_0 + \vs{x}t_1 \right)^2} \times e^{-\alpha \left(y_1 - \vp{y}\tau \right)^2 -\alpha \left(y_2 - \vp{y}\tau \right)^2} \nonumber \\ 
&& \times\> e^{-\beta(t_2 - t_1)\left(y_2 - y_1 + \vs{y}(t_2 - t_1)\right)^2 -\beta(t_1) \left( y_1 + \vs{y}t_1 \right)^2} \nonumber \\
&& \times\> e^{-\alpha \left(z_1 - \vp{z}\tau \right)^2 -\alpha \left(z_2 - \vp{z}\tau \right)^2 -\beta(t_2 - t_1)\left(z_2 - z_1 + \vs{z}(t_2 - t_1)\right)^2} \nonumber \\ 
&& \times\> e^{ -\beta(t_1) \left( z_1 + \vs{z}t_1 \right)^2} \dif x_1 \dif x_2 \dif y_1 \dif y_2 \dif z_1 \dif z_2.   	 
\end{IEEEeqnarray} 
To solve the multiple integrals in \eqref{Eq.ACF_Proof2}, we use the PDF integration formula for multivariate Gaussian distributions. In particular, let us assume that vector $\mathbf{X} = [x_1,y_1,z_1,x_2,y_2,z_2]^{\intercal}$ has a multivariate Gaussian distribution with mean vector $\bm{\mu} = \mathcal{E}\{ \mathbf{X} \} \in \mathbb{R}^6$ and covariance matrix $\bm{\Sigma} = \mathcal{E} \{(\mathbf{X} - \bm{\mu})(\mathbf{X} - \bm{\mu})^\intercal \}$. Then, the well-known PDF of $\mathbf{X}$ is given by 
\begin{IEEEeqnarray}{rCl}
	\label{Eq.MultivariateGaussian}
\hspace{-5 mm} \textit{{\large f}}_{\mathbf{X}}(x_1,y_1,z_1,x_2,y_2,z_2) = \frac{\exp \left( -\frac{1}{2}  (\mathbf{X} - \bm{\mu})^\intercal \bm{\Sigma}^{-1} (\mathbf{X} - \bm{\mu})\right)}{(2\pi)^3 \sqrt{\text{det}\left( \bm{\Sigma} \right)}},
\end{IEEEeqnarray}
where $\text{det}(\cdot)$ denotes the determinant. It can be easily verified that for mean vector $\bm{\mu} = [\mu_{x_1}, \mu_{y_1}, \mu_{z_1}, \mu_{x_2}, \mu_{y_2}, \mu_{z_2}]$ and inverse covariance matrix 
\begin{IEEEeqnarray}{C}
	\label{Eq.InverseCovMat}
	\Sigma^{-1} = \left[ \begin{matrix}
     \vartheta & 0 & 0 & \psi & 0 & 0 \\
     0 & \vartheta & 0 & 0 & \psi & 0 \\
     0 & 0 & \vartheta & 0 & 0 & \psi \\
     \psi & 0 & 0 & \varepsilon & 0 & 0 \\
     0 & \psi & 0 & 0 & \varepsilon & 0 \\
    0 & 0 & \psi & 0 & 0 & \varepsilon 
\end{matrix} \right],
\end{IEEEeqnarray}
where $\mu_{x_1}, \mu_{y_1}, \mu_{z_1}, \mu_{x_2}, \mu_{y_2}, \mu_{z_2}, \vartheta, \varepsilon$, and $\psi$ are given in \eqref{Eq.Param} (on the top of the page), 
$\exp \left( -\frac{1}{2}  (\mathbf{X} - \bm{\mu})^\intercal \bm{\Sigma}^{-1} (\mathbf{X} - \bm{\mu})\right)\times\exp(\kappa_x + \kappa_y + \kappa_z) $ (with $\kappa_x, \kappa_y, \kappa_z$ as given in \eqref{Eq.ACF_kappa}) is equal to the integrands in \eqref{Eq.ACF_Proof2}. Now, given that $\int_{-\infty}^{+\infty} \cdots \int_{-\infty}^{+\infty} \textit{{\large f}}_{\mathbf{X}}(x_1,y_1,z_1,x_2,y_2,z_2) \dif x_1 \ldots \dif z_2 = 1$, $\phi(t_1, t_2)$ can be written as  
\begin{IEEEeqnarray}{C}
	\label{Eq.ACFProof3} 
	\phi(t_1,t_2) = \varphi^2 \lambda(t_2 - t_1) \lambda(t_1) \exp(\kappa_x + \kappa_y + \kappa_z) (2\pi)^3 \sqrt{\text{det}(\bm{\Sigma)}}. \nonumber \\*
\end{IEEEeqnarray}
Given $\bm{\Sigma}^{-1}$ in \eqref{Eq.InverseCovMat}, after some calculations, it can be shown that  
\begin{IEEEeqnarray}{C}
	\label{Eq.DeterminantSol}
	\text{det}\left( \bm{\Sigma} \right) = \frac{1}{\left(\vartheta \times \varepsilon - \psi^2 \right)^3}.
\end{IEEEeqnarray}
Finally, substituting \eqref{Eq.DeterminantSol} into \eqref{Eq.ACFProof3} leads to \eqref{Eq.ACF}.

\section{Proof of Theorem 3}
\label{App.2}
For calculation of $\textit{{\large F}}_{h(t,\tau)}(h)$, we first find the distribution of $| \vec{r}(t) - \vecp \tau |^2$. Given the PDF of random variable $\vec{r}(t)$ in \eqref{Eq.PDFr(t)}, we obtain for the elements of the vector $\vec{r}(t) - \vecp \tau = [X(t), Y(t), Z(t)]$ 
\begin{IEEEeqnarray}{rCl}
	\label{Eq.equivalentr(t)_Pre} 
	X(t) & \sim & \mathcal{N}\left(x_0 - \vs{x}t - \vp{x}\tau, 2D_2 t\right), \nonumber \\
	Y(t) & \sim & \mathcal{N}\left(- \vs{y}t - \vp{y}\tau, 2D_2 t\right), \nonumber \\
	Z(t) & \sim & \mathcal{N}\left(- \vs{z}t - \vp{z}\tau, 2D_2 t\right). 
\end{IEEEeqnarray} 
We can rewrite $|\vec{r}(t) - \vecp \tau|^2 $ as follows
\begin{IEEEeqnarray}{rCl}
	\label{Eq.equivalentr(t)}
	|\vec{r}(t) - \vecp \tau|^2 & = & X^2(t) + Y^2(t) + Z^2(t) \nonumber \\
	 & = & 2D_2t \times \left( \tilde{X}^2(t) + \tilde{Y}^2(t) + \tilde{Z}^2(t) \right)\nonumber \\ 	 & = & 2D_2t \times  \tilde{r}^2(t),
\end{IEEEeqnarray}
where 
\begin{IEEEeqnarray}{rCl} 
\tilde{X}(t) & \sim & \mathcal{N}\left((x_0 - \vs{x}t - \vp{x}\tau)/\sqrt{(2D_2t)}, 1\right), \nonumber \\ 
 \tilde{Y}(t) & \sim & \mathcal{N}\left((-\vs{y}t - \vp{y}\tau)/\sqrt{(2D_2t)},1\right), \nonumber \\  \tilde{Z}(t)& \sim & \mathcal{N}\left((- \vs{z}t - \vp{z}\tau)/\sqrt{(2D_2t)},1\right). 
\end{IEEEeqnarray}
Given \eqref{Eq.equivalentr(t)}, we can rewrite the CIR in \eqref{Eq.CIRTimVar} as $h(t,\tau) = \varphi \exp(-2$ $\times D_2t\alpha \tilde{r}^2(t))$, where $\tilde{r}^2(t)$ follows a noncentral chi-square distribution with $k = 3$ degrees of freedom and noncentrality parameter $\gamma(t) = |\vec{r}_0 - \vecs t - \vecp \tau|^2 / 2D_2 t$, i.e., $\tilde{r}^2(t) \sim \chi_3^2(\gamma(t))$. Therefore, we can calculate the CDF of the CIR of the mobile MC channel as follows
\begin{IEEEeqnarray}{rCl}
	\label{Eq.CIRDerivation}
	\textit{{\large F}}_{h(t,\tau)}(h) & = & \pr \left( h(t,\tau) < h \right) \nonumber \\ 
	& = &  \pr \left( \varphi \exp\left( - 2D_2t\alpha \tilde{r}^2(t) \right) \leq h \right) \nonumber \\
	& = & \pr \left( \tilde{r}^2(t) \geq \frac{\ln \left( \varphi / h \right)}{2D_2t \alpha} \right) \nonumber \\ 
	& = & 1 - \pr \left( \tilde{r}^2(t) < \frac{\ln \left( \varphi / h \right)}{2D_2t \alpha} \right),
\end{IEEEeqnarray}
where $\ln(\cdot)$ denotes the natural logarithm. The last term on the right-hand side of \eqref{Eq.CIRDerivation} is the CDF of random variable $\tilde{r}^2(t)$. The CDF of a random variable $U \sim \chi_{k}^2(\gamma)$, i.e., $\pr (U \leq u)$, is given by $1 - Q_{k/2}(\sqrt{\gamma}, \sqrt{u})$, where $Q_m(a,b)$ denotes the generalized Marcum Q-function of order $m$ \cite[Eq.~(4.33)]{SimAlo}
\begin{IEEEeqnarray}{C}
	\label{Eq.MarcumQFun}
\hspace{-3mm}	Q_m(a,b) = \frac{1}{a^{m-1}} \int_{b}^{\infty} x^m \exp \left( - \frac{x^2 + a^2}{2} \right) I_{m-1}(ax) \dif x,
\end{IEEEeqnarray}
where $I_{m}(\cdot)$ is the $m$th-order modified Bessel function of the first kind. Given \eqref{Eq.CIRDerivation} and \eqref{Eq.MarcumQFun}, we obtain 
\begin{IEEEeqnarray}{C}
	\label{Eq.CDFofCIR}
	\textit{{\large F}}_{h(t,\tau)}(h) = Q_{3/2}\left( \frac{r^{\text{eq}}(t)}{\sqrt{2D_2t}}, \sqrt{\frac{\ln \left( \varphi / h \right)}{2D_2t \alpha}} \right). 
\end{IEEEeqnarray}
In order to further simplify the expression derived in \eqref{Eq.CDFofCIR}, we use the closed-form representation of $Q_m(a,b)$ proposed in \cite{LiKam}. There, it has been shown that for the case of $m = 0.5n$, where $n$ is an \emph{odd} positive integer, $Q_m(a,b)$ is given by \cite[Eq.~(11)]{LiKam}
\begin{IEEEeqnarray}{rCl}
	\label{Eq.MarcumQFunAlter}
	Q_m(a,b) & = & \frac{1}{2}\erfc \left( \frac{a + b}{\sqrt{2}} \right) + \frac{1}{2} \erfc \left( \frac{b - a}{\sqrt{2}} \right) \nonumber \\ 
	&& +\> \frac{1}{a\sqrt{2\pi}} \sum_{k=0}^{m-1.5} \frac{b^{2k}}{2^k} \sum_{q = 0}^{k} \frac{(-1)^q (2q)!}{(k-q)!q!} \nonumber \\
	&& \> \times \left\lbrace \sum_{i = 0}^{2q} \frac{1}{(ab)^{2q-i} i!} \left[ (-1)^i \exp \left( -\frac{(b-a)^2}{2} \right) \right. \right. \nonumber \\ 
	&& \left. \left. -\> \exp \left( -\frac{(b+a)^2}{2} \right) \right] \right\rbrace,a>0, b \geq 0.
\end{IEEEeqnarray}
After substituting $m = 3/2 = 0.5\times3$, $a = r^{\text{eq}}(t)/\sqrt{2D_2t}$, and $b = \sqrt{\ln(\varphi/h)/2D_2t\alpha}$ from \eqref{Eq.CDFofCIR} into \eqref{Eq.MarcumQFunAlter}, $\textit{{\large F}}_{h(t,\tau)}(h)$ simplifies to \eqref{Eq.CDFofCIRSimplified}.
\bibliographystyle{IEEEtran}
\bibliography{IEEEabrv,Library}
\begin{IEEEbiography}[{\includegraphics[width=1in,height=1.25in,clip,keepaspectratio]{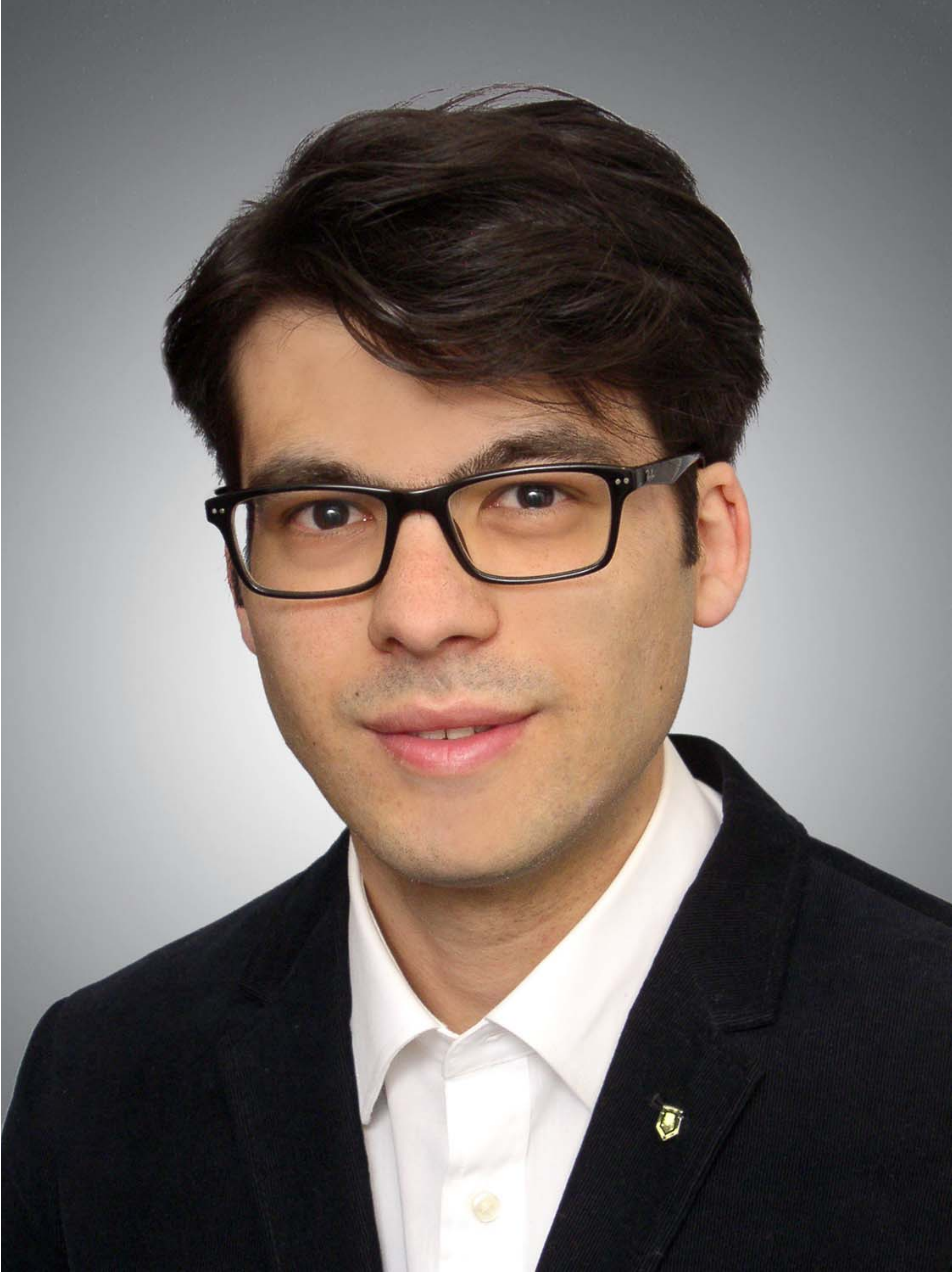}}]{Arman Ahmadzadeh} (S'14) received the B.Sc. degree in electrical engineering from the Ferdowsi University of Mashhad, Mashhad, Iran, in 2010, and the M.Sc. degree in communications and multimedia engineering from the Friedrich-Alexander University, Erlangen, Germany, in 2013, where he is currently pursuing the Ph.D. degree in electrical engineering with the Institute for Digital Communications. His research interests include physical layer molecular communications. Arman served as a member of Technical Program Committees of the Communication Theory Symposium for the IEEE International Conference on Communications (ICC) 2017 and 2018. Arman received several awards including the ``Best Paper Award'' from the IEEE ICC in 2016, ``Student Travel Grants'' for attending the Global Communications Conference (GLOBECOM) in 2017, and was recognized as an Exemplary Reviewer of the \textsc{IEEE Communications Letters} in 2016.     
\end{IEEEbiography}

\begin{IEEEbiography}[{\includegraphics[width=1in,height=1.25in,clip,keepaspectratio]{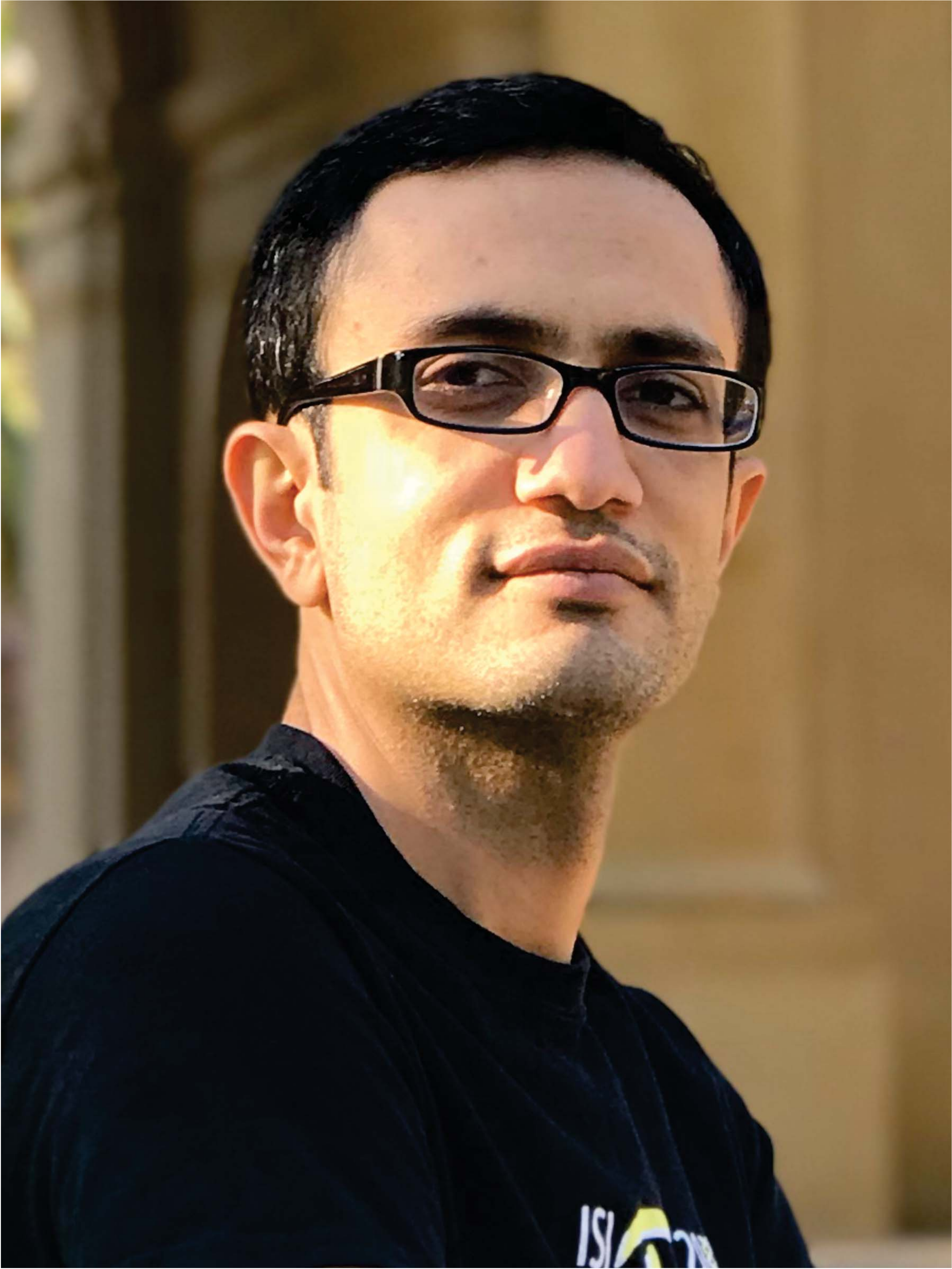}}]{Vahid Jamali} (S'12) received the B.S. and M.S. degrees (Hons.) in electrical engineering from the K. N. Toosi University of Technology, Iran, in 2010 and 2012, respectively. He is working toward his Ph.D. degree at the Friedrich-Alexander University (FAU) of Erlangen-Nuremberg, Germany. In 2017, he was a visiting research scholar at the Stanford University, USA. His research interests include wireless communications, molecular communications, multiuser information theory, and signal processing. He served as a member of Technical Program Committees for several conferences including the IEEE PIMRC, IEEE VTC, IEEE Int. BlackSeaCom, and ICNC. Vahid has received several awards including the ``Winner of the Best 3 Minutes Ph.D. Thesis (3MT) Presentation'' from the IEEE WCNC in 2018,``Doctoral Scholarship'' from the German Academic Exchange Service (DAAD) in 2017, the ``Best Paper Award'' from IEEE ICC in 2016, ``Student Travel Grants'' for attending the SP Coding and Information School, Sao Paulo, Brazil in 2015, the Training School on Optical Wireless Communications, Istanbul, Turkey in 2015, and the IEEE ICC, Paris, France in 2017, and ``Exemplary Reviewer Certificates'' from the \textsc{IEEE Communications Letters} in 2014 and \textsc{IEEE Transactions on Communications} in 2017.
\end{IEEEbiography}
\begin{IEEEbiography}[{\includegraphics[width=1in,height=1.25in,clip,keepaspectratio]{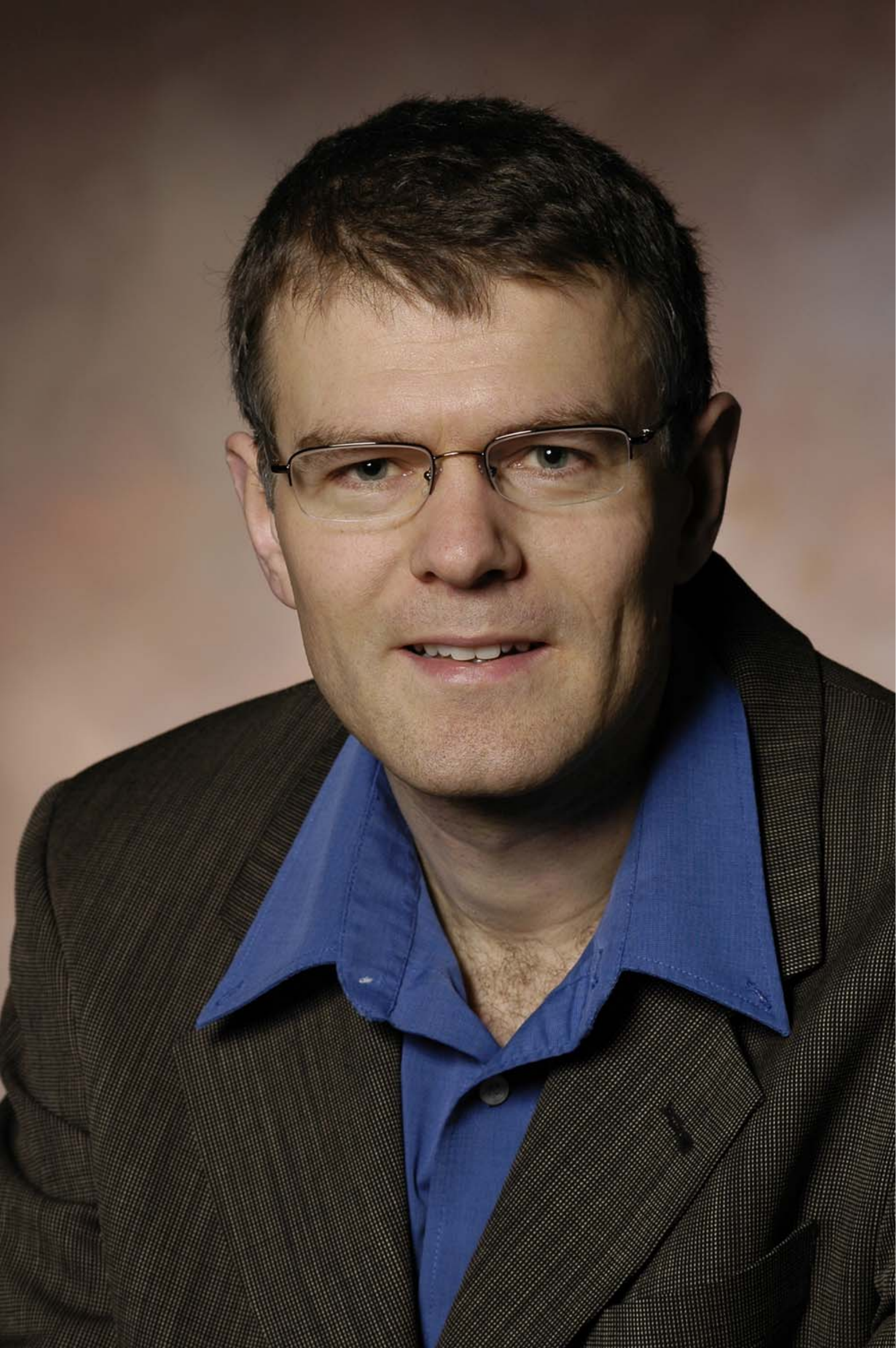}}]{Robert Schober}(S'98, M'01, SM'08, F'10) received the Diplom (Univ.) and the Ph.D. degrees in electrical engineering from the Friedrich-Alexander University of Erlangen-Nuremberg (FAU), Germany, in 1997 and 2000, respectively. From 2002 to 2011, he was a Professor and Canada Research Chair at the University of British Columbia (UBC), Vancouver, Canada.
Since January 2012 he is an Alexander von Humboldt Professor and the Chair for Digital Communication at FAU. His research interests fall into the broad areas of Communication Theory, Wireless Communications, and Statistical Signal Processing.

Robert received several awards for his work including the 2002 Heinz Maier-Leibnitz Award of the German Science Foundation (DFG), the 2004 Innovations Award of the Vodafone Foundation for Research in Mobile Communications, a 2006 UBC Killam Research Prize, a 2007 Wilhelm Friedrich Bessel Research Award of the Alexander von Humboldt Foundation, the 2008 Charles McDowell Award for Excellence in Research from UBC, a 2011 Alexander von Humboldt Professorship, a 2012 NSERC E.W.R. Stacie Fellowship, and a 2017 Wireless Communications Recognition Award by the IEEE Wireless Communications Technical Committee. He is listed as a 2017 Highly Cited Researcher by the Web of Science and a Distinguished Lecturer of the IEEE Communications Society (ComSoc). Robert is a Fellow of the Canadian Academy of Engineering and a Fellow of the Engineering Institute of Canada. From 2012 to 2015, he served as Editor-in-Chief of the \textsc{IEEE Transactions on Communications}. Currently, he is the Chair of the Steering Committee of the \textsc{IEEE Transactions on Molecular, Biological and Multiscale Communication}, a Member of the Editorial Board of the Proceedings of the IEEE, a Member at Large of the Board of Governors of ComSoc, and the ComSoc Director of Journals.
\end{IEEEbiography}
 
\end{document}